\documentclass[11pt]{article}
\usepackage{jheppub}

\usepackage{amsmath,graphicx,amssymb,subfigure}
\usepackage[all]{xy}

\newcommand{\be}{\begin{equation}}
\newcommand{\ee}{\end{equation}}
\newcommand{\beqa}{\begin{eqnarray}}
\newcommand{\eeqa}{\end{eqnarray}}

\def\dd{\mathrm{d}}

\def\cM{{\cal M}}
\def\cO{{\cal O}}

\title{D3/D7 Quark-Gluon Plasma with Magnetically Induced Anisotropy}

\author{Martin Ammon$^1$,}
\author{Veselin Filev$^2$,}
\author{Javier Tarr\'\i o$^3$,}
\author{Dimitrios Zoakos$^4$}

\affiliation{$^1$ Department of Physics and Astronomy, University of California,\\
Los Angeles, CA 90095, United States}
\affiliation{$^2$ School of Theoretical Physics, Dublin Institute for Advanced Studies,\\ 
10 Burlington Road, Dublin 4, Ireland}

\affiliation{$^3$ Institute for Theoretical Physics and Spinoza Institute, Universiteit Utrecht,\\ 
3584 CE, Utrecht, The Netherlands}

\affiliation{$^4$ Centro de F\'\i sica do Porto and Departamento de F\'\i sica e Astronomia,\\ Faculdade de Ci\^encias da Universidade do Porto,\\ 
Rua do Campo Alegre 687, 4169--007 Porto, Portugal}
\emailAdd{ammon@physics.ucla.edu} 
\emailAdd{vfilev@stp.dias.ie} 
\emailAdd{l.j.tarriobarreiro@uu.nl} 
\emailAdd{dimitrios.zoakos@fc.up.pt}

\abstract{We study the effects of the temperature and of a magnetic field in the setup of an intersection of D3/D7 branes, where a large number of D7 branes is smeared in the transverse directions to allow for a perturbative solution in a backreaction parameter. The magnetic field sources an anisotropy in the plasma, and we investigate its physical consequences for the thermodynamics and energy loss of particles probing the system. In particular we comment on the stress-energy tensor of the plasma, the propagation of sound in the directions parallel and orthogonal to the magnetic field, 
the drag force of a quark moving through the medium and jet quenching.}


\subheader{ITF-UU-12/23, SPIN-12/21, DIAS-STP-12-04}

\begin{document}

\maketitle


\section{Introduction}

One of the amazing developments emerging from the research in string theory, is the idea of a gauge/gravity correspondence  \cite{Maldacena:1997re}. 
The remarkable feature of this correspondence is that it relates the strongly coupled regime of the gauge theory to the weakly coupled regime of the string theory and vice-versa. Consequently, it has become a powerful tool in studying strongly interacting systems by using a conjectured dual weakly coupled string/gravitational theory.  
At present, holographic descriptions of non-perturbative phenomena include, among others
applications to condensed matter physics,  high energy physics and quark-gluon plasma. 

\paragraph{}

One of the most distinctive uses of the gauge/gravity correspondence has been the study of the physics of  heavy ion collisions. Through collisions at Brookhaven and LHC a strongly coupled plasma of quarks and gluons was created which cannot be described by the standard perturbative techniques. Also other methods such as Lattice Gauge theory fail in computing transport coefficients of the plasma and the rapid thermalization rate of the quark gluon plasma observed. This is where the gauge/gravity duality enters in the field and provides interesting new insights. For example, for large-$N$ gauge theories at strong coupling, gauge/gravity duality predicts that the ratio of the shear viscosity to entropy density is $1/4\pi$ (see \cite{Policastro:2001yc}) in natural units and therefore very close to the measured value. The small value of the ratio of the shear viscosity to entropy density can be understood as an effect of the strong coupling of the system. Within the framework of Gauge/Gravity duality we can also compute the thermalization rate  $\tau_{th}\sim 0.5 \text{fm}$ of the plasma \cite{Chesler:2009cy,Balasubramanian:2011ur}, which is in  agreement with the observed value indicating again the strong coupling nature of the plasma. The energy loss of the heavy/energetic partons in the plasma also acquires a gravity dual description (see \cite{Dragforce}).

\paragraph{}

Despite the remarkable insights into the quark-gluon plasma and QCD in general gained by studying gauge/gravity dualities, the application of the correspondence to real-world systems such as QCD remains a challenge and has to be developed further. So far, we do not have a rigorous string dual of QCD at hand. However, under extreme external parameters (such as temperature and chemical potential)
different gauge theories exhibit similar properties. Therefore,  it is natural to apply holographic techniques to study phenomena which are believed to be of universal nature.

\paragraph{}

An important example in this class of phenomena is the effect of mass generation and spontaneous chiral symmetry breaking in the presence of  an external magnetic field. 
The effect is known as magnetic catalysis and has been shown insensitive to the microscopic physics underlying the low energy effective theory. 
Using conventional field theory methods, the magnetic catalysis has been demonstrated in various (1+2) and (1+3)-dimensional field theories  \cite{Catalysis-FT}, 
while the  holographic study of the effect initiated in \cite{Filev:2007gb}\footnote{For a comprehensive review we refer the reader to \cite{Filev:2010pm}.}. 
Additional holographic studies of magnetic catalysis at finite temperature or chemical potential appear in \cite{Catalysis-AdS}. 

\paragraph{}

Until recently all the holographic studies of the magnetic catalysis were in the probe approximation, where 
the backreaction of the flavor branes on the supergravity background is neglected \cite{Karch:2002sh}. 
On the field theory side, this corresponds to an approximation in which the
flavor degrees of freedom $N_f$ are much smaller than the color ones $N_c$. 
Unquenching the holographic description means a large number of flavor branes that backreact on the geometry.
Due to the technical difficulties that arise from a set of localized flavor branes, we distribute them along the compact directions \cite{Bigazzi:2005md}. 
This procedure is called smearing\footnote{For a detailed review on the smearing see the review \cite{arXiv:1002.1088}, while for other solutions employing this
technique that appeared after the review see \cite{allsusyunquenched}.} and restores a significant part of the global symmetry of the geometry. 

\paragraph{}

A promising framework for the construction of such a geometry was started in \cite{Benini:2006hh}, where the ten-dimensional supergravity solutions including the backreaction of a large number of D7-branes in $AdS_5\times X_5$ (with $X_5$ any squashed Sasaki-Einstein manifold) was introduced. This was further developed in \cite{Bigazzi:2009bk}, where the  black-hole solution dual
to the non-conformal plasma of flavored ${\cal N} = 4$ supersymmetric Yang-Mills theory is presented\footnote{All the hydrodynamic transport coefficients of the model were analyzed in \cite{hydro}, while the addition of a finite baryon density was presented in \cite{Bigazzi:2011it}. 
For a review on unquenching the Quark Gluon Plasma see \cite{Bigazzi:2011db}. }. 
The authors outline the smearing procedure, 
derive the corresponding equations of motion and present a perturbative solution for general massless non-supersymmetric flavor D7--brane embeddings. 

\paragraph{}

The first steps towards unquenching the holographic description of magnetic catalysis have been undertaken in \cite{Filev:2011mt} and \cite{ Erdmenger:2011bw}. 
More specifically in \cite{Filev:2011mt}, a string dual to SU($N_c$) ${\cal N} = 4$ SYM coupled to $N_f$ \emph{massless} fundamental flavors 
in the presence of an external magnetic field is presented. For sufficiently strong magnetic field, the supergravity background is unstable, suggesting that the theory undergoes a phase transition to a stable phase with dynamically generated mass for the matter fields. In \cite{ Erdmenger:2011bw}, the external magnetic field couples to
 $N_f$ \emph{massive} fundamental flavors and the background has a hollow cavity in the bulk of the geometry, where it is similar to the supergravity dual of a 
${\cal N}=1 $ non-commutative SYM. The radius of this cavity is related to the dynamically generated mass of the fundamental fields.
After developing an appropriate renormalization scheme, the free energy and the condensate can be expanded in powers of the perturbative parameter. 
While at leading order, both agree with the previously obtained results in the probe approximation, 
at next to leading order the effect of magnetic catalysis is enhanced and the contribution to the condensate runs logarithmically with the finite cutoff  $\Lambda_{UV}$.

\paragraph{}


An overview of the paper is as follows: In section
\ref{sec:BH} we continue the studies initiated in \cite{Filev:2011mt, Erdmenger:2011bw} and present
a string dual to the finite temperature SU($N_c$) ${\cal N} = 4$ SYM
coupled to $N_f$ \emph{massless} fundamental matter
in the presence of an external magnetic field. The solution is
analytic and perturbative in a parameter that counts the number of
internal fundamental loops.
Given the non illuminating expressions for the functions of the
background we provide some numerical plots, and since we have a
perturbative solution we supplement it with a
hierarchy of scales.

\paragraph{}

In section \ref{sec:TD} we study the thermodynamics of the anisotropic
black hole, which provides a non trivial check for the validity of the
gravity solution.
Since the solution is first order in the expansion parameter, our
computations have some overlap with those of \cite{Albash:2007bk} and
extend those of
\cite{Bigazzi:2009bk} in the presence of an external magnetic field.
While in the absence of a magnetic field the breaking of conformal
invariance happens at second order
in the expansion parameter \cite{Mateos:2007vn,
Bigazzi:2009bk,Bigazzi:2011it}, in its presence conformal invariance
breaks at first order.

\paragraph{}

In section \ref{sec:setensor} we holographically calculate the stress
energy tensor of the boundary field theory. The presence of the
magnetic field sources an anisotropy in the medium,
which is realized through a difference between the pressure transverse
to the magnetic field and the pressure along the direction of the
magnetic field.
We present thermodynamic arguments supporting the holographic computation.

\paragraph{}

In section \ref{sec:loss} we calculate the energy loss of the partons
as they propagate through the anisotropic plasma.
The jet quenching parameter depends on the relative orientation
between the anisotropic direction, the direction of motion
of the parton and the direction along which the momentum broadening is measured.
We consider a parton moving parallel to the magnetic field with the
momentum broadening taking place in the transverse plane. The presence
of the
magnetic field enhances or reduces the jet quenching parameter of a
theory without magnetic field, depending on the conditions we use to make the
comparison.
The drag force experienced by an infinitely massive quark propagating
at a general angle through the plasma is calculated using an
appropriate set up
to compensate the Lorentz force on the probe quark.  In this way we
obtain an expression reflecting the anisotropy of the plasma due to
the external magnetic field.


\section{Constructing the black hole}\label{sec:BH}

The present section is devoted to the construction of a supergravity background describing an anisotropic black hole.
The field theory duals are realized on the intersection between a set of $N_c$ {\it color} D3-branes and a 
set of $N_f$, homogeneously smeared, {\it flavor} D7--branes, with an additional coupling between the fundamental fields and an external magnetic field.


\subsection{Setup} \label{sec:setup}

The smearing of the flavor D7-branes  allows for an ansatz where all the functions of the background depend just on the radial coordinate. 
Having this in mind and inspired by \cite{Bigazzi:2009bk, Filev:2011mt, Erdmenger:2011bw}, we adopt the following ansatz for the metric
\begin{eqnarray} \label{10dmetric}
ds_{10}^2 & = & h^{-\frac{1}{2}}\, \Big[- \, b_T^2 \, dt^2 \, + \,  b \, \left(dx_1^2 \, + \, dx_2^2\right) \, + \,  dx_3^2 \, \Big] 
\nonumber  \\ 
&&  \qquad \quad \qquad \quad
+ \, \,  h^\frac{1}{2} \, 
\Big[\, b^2 \,  b_T^2 \,  S^8 \, F^2 \, d\sigma^2 \, +  \, S^2 \, ds_{CP^2}^2 \, + \,  F^2 \left(d\tau \, + \,  A_{CP^2} \right)^2  \, \Big] \ ,
\end{eqnarray}
where the $CP^2$ metric is given by 
\begin{eqnarray}
ds_{CP^2}^2&=&\frac{1}{4} d\chi^2+ \frac{1}{4} \cos^2 \frac{\chi}{2} (d\theta^2 +
\sin^2 \theta d\varphi^2) + \frac{1}{4} \cos^2 \frac{\chi}{2} \sin^2 \frac{\chi}{2}(d\psi + \cos \theta d\varphi)^2
\ , \nonumber \\
A_{CP^2}&=& \frac12\cos^2 \frac{\chi}{2}(d\psi + \cos \theta d\varphi)\,\,.
\label{cp2metric}
\end{eqnarray}
The range of the angles is $0\leq (\chi, \theta) \leq \pi$,  $0\leq (\varphi, \tau) < 2\pi$, $0\leq \psi< 4 \pi$.
The ansatz for the NSNS and the RR field strengths is given by
\begin{eqnarray} \label{NS+RR}
& B_{2}  =  H dx^1\wedge dx^2 \ , \quad
C_{2} = J \,dt \wedge dx^3 \,,  & 
\nonumber\\
& F_{5}  =  Q_c\,(1\,+\,*)\varepsilon(S^5)\ , \quad
F_{1} = Q_f \,(d\tau + A_{CP^2})\ , \quad
F_{3} = d C_2\,+\, B_{2} \wedge F_1 \ , &
\end{eqnarray}
where $\varepsilon(S_5)$ is the volume element of the internal space\footnote{With $\int \, \varepsilon(S_5) \, = \, \text{Vol}(S^5) \, = \, \pi^3$.} 
and $Q_{c}, Q_{f}$ are related to
the number of different colors and flavors in the following way
\begin{equation}
N_c = \frac{Q_c\, Vol(X_{SE})}{(2\pi)^4g_s \,\alpha'^2} \ , \qquad
N_f = \frac{4\,Q_f\,Vol(X_{SE})}{Vol(X_3) g_s} \ .\label{QcQf}
\end{equation}
In our case $X_{SE}=S^5$ and the $X_3=S^3$, a 3-sphere with volume $2\pi^2$.
The fact that the flavors are massless is encoded in the independence of $F_1$ on $\sigma$, see \cite{Benini:2006hh, Bigazzi:2008zt} . 
All the functions that appear in the ansatz, $h, \, b_T,\, b,\, S,\, F,\, \Phi, J$ and $H$,
depend on the radial variable $\sigma$ only. In the convention we follow, $S$  and  $F$ have dimensions of length, 
$b,\, b_T,\, h,\, J$ and $H$ are dimensionless and $\sigma$ has a dimension of length${}^{-4}$.
The function $b$ in the ansatz for the metric reflects the breaking of the $SO(1,3)$ Lorentz symmetry down to $SO(1,1)\times SO(2)$. 
The blackening function $b_T$ allows for the existence of solutions with a black brane, whose horizon sits at a position 
$\sigma_h$ such that $b_T(\sigma_h)=0$, and which allows to study the field theory at finite temperature.

Solving the 10d equation of motion for $F_3$, we need to impose the following relation
\begin{equation} \label{10d-F3}
J'=Q_c\frac{e^{-\Phi} b_T^2}{h} (H-H_{0}) \,  ,
\end{equation}
where $H_{0}$ is an integration constant. In the next subsection we will keep the function $J$ and will see how this relation appears from an effective one-dimensional Lagrangian.


\subsection{Effective actions and equations of motion}

The action for the Type IIB supergravity plus the contribution from the $N_f$ D7--branes in the Einstein frame is 
\begin{equation}
S=S_{IIB} + S_{fl} \ , \label{genact}
\end{equation}
where the relevant terms of the $S_{IIB}$ action are
\begin{eqnarray} \label{TypeIIB action}
S_{IIB}&=&\frac{1}{2\kappa_{10}^2}\int d^{10} x \sqrt{-g} \Bigg[ R
- {\frac{1}{2}} \partial_M\Phi \partial^M\Phi
- \frac{1}{2} e^{2\Phi}F_{(1)}^2 
- \frac{1}{2} \frac{1}{3!} e^{\Phi}  F_{(3)}^2
- \frac{1}{2} \frac{1}{5!} F_{(5)}^2   \\
&&\qquad \qquad \qquad \qquad \qquad  \quad \qquad \qquad  \,
- \frac{1}{2} \frac{1}{3!} e^{-\Phi}  H_{(3)}^2\Bigg] 
- \frac{1}{2\kappa_{10}^2}\, \int C_4\wedge H_3\wedge F_{3} \ , \nonumber
\end{eqnarray}
and the action for the flavor D7--branes takes the usual DBI+WZ form
\begin{equation}
S_{fl} = -T_7 \sum_{N_f} \Bigg[ \int d^8x\, e^\Phi 
\sqrt{-\det (\hat{G}+e^{-\Phi/2} {\cal F}}) \, 
- \, \int \left(\hat{C}_8 + \hat{C}_6 \wedge  \cal F \right) \Bigg]\,,
\label{actionflav}
\end{equation}
with ${\cal F} \equiv  \hat B_2 + 2 \pi \alpha' F$. In those expressions $B_2$ denotes a non-constant NSNS potential which will model the magnetic field,
$F$ the worldvolume gauge field and the hat refers to the pullback of the quantities, 
along the worldvolume directions of the D7--brane.  The gravitational constant and D7--brane tension, 
in terms of string parameters, are
\begin{equation}
\frac{1}{2\kappa_{10}^2} = \frac{T_7}{g_s} = \frac{1}{(2\pi)^7g_s^2 \alpha'^4} \ .
\end{equation}
We plug our ansatze, \eqref{10dmetric} and \eqref{NS+RR}, into \eqref{genact} and integrate out all the directions except the radial one, since the dependence is trivial. 
After an integration by parts to get rid of second derivatives we obtain the following expression
\begin{equation} \label{eff-action}
S_{eff}=\frac{\pi ^3 V_{1,3}}{2\kappa_{10}^2}\int {\cal L}_{eff} \, d\sigma
\end{equation}
where $V_{1,3}$ is the volume of the Minkowski space and the one-dimensional effective lagrangian ${\cal L}_{eff} $ is given appendix  \ref{eff-EOM}. 
Since the function $J$ enters in the effective action only via its radial derivative, there is a first integration given by a conserved quantity. 
We fix this constant of motion in the following way
\begin{equation} \label{defJ}
\frac{\partial {\cal L}_{eff}}{\partial J'} \, \equiv - \, Q_c H_{0} \quad \Rightarrow \quad 
J' \,=\, \frac{e^{-\Phi}\,Q_c\, b_T^2}{h} \left(H-\, H_{0}\right)\ .
\end{equation}
which is precisely \eqref{10d-F3}. 
The next step is to use \eqref{defJ} to eliminate $J'$ in favor of $H$ in \eqref{eff-action}, after performing the following 
Legendre transformation 
\begin{equation}  \label{L-effective2}
\tilde {\cal L}_{eff} = {\cal L}_{eff}-\frac{\delta {\cal L}_{eff}}{\delta J'} \, J'\Bigg|_{J' \equiv J'(H)} \ ,
\end{equation}
and then calculate the Euler-Lagrange equations from the transformed action \eqref{L-effective2}. The equations of motion are given in appendix  \ref{eff-EOM}.

Setting $Q_f=0$ in the transformed action, the Euler-Lagrange equations imply that a solution with $H\neq0$ is given by (black) $AdS_5\times X_{SE}$ with $\Phi=\Phi_*$ and $H=H_{0}$ constants. We will use this solution later on as a starting point to obtain a black brane solution with backreacted flavor in the presence of a non trivial $H$.
 
 It is worth noting that by demanding   $\partial_{ J' } {\cal L}_{eff}  =  -  Q_c H_{0}$ exactly, with $H_{0}$ the value of the magnetic field in the unflavored limit, we are enforcing the field $J$ to vanish when $N_f\to0$. As such, $J$ reflects  magnetic effects by providing a field connected holographically to the magnetization of the system, as we will see.

The equation for the blackening factor \eqref{diff-bT} decouples from the rest and can be solved analytically
\begin{equation}
b_T^2 = e^{-4r_h^4 \sigma} \ ,
\end{equation}
where $r_h$ is a non-extremality parameter coming from the integration constants. 
The position of the horizon is at $\sigma\to\infty$, whereas the boundary would be at $\sigma=0$ (there is an additional integration constant corresponding to a shift in $\sigma$, which we set to zero).


\paragraph{Reduced five-dimensional action}

For the calculation of the stress-energy tensor in section \ref{sec:setensor}, we find convenient to write as well  a truncated five-dimensional action, obtained after integrating out the compact Sasaki-Einstein manifold  in \eqref{10dmetric}. Denoting the effective metric as $g_{\mu\nu}$, the action is
\be\label{5daction}
S_{5d}  = \frac{1}{2\kappa_5^2} \int \dd^5x \sqrt{-g}\left[ {\cal L}_{kin} + {\cal L}_{pot}\right] + \frac{1}{2\kappa_5^2} \int \dd^5x {\cal L}_{top}   \ ,
\ee
where the kinetic, potential and topological terms are given by
\begin{eqnarray}
{\cal L}_{kin} & = & R[g] - \frac{40}{3} \partial_\mu f \partial^\mu f - 20 \partial_\mu w \partial^\mu w - \frac{1}{2} \partial_\mu \Phi \partial^\mu \Phi - \frac{1}{12} e^{\Phi - \frac{20}{3}f} F_{\mu\nu\rho}F^{\mu\nu\rho} 
 \\
&- & \frac{1}{12} e^{-\Phi - \frac{20}{3}f} H_{\mu\nu\rho}H^{\mu\nu\rho}   \ , \nonumber
\\
{\cal L}_{pot} & = & \, - \, 4e^{\frac{16}{3}f+2w}\left( e^{10w}-6 \right) \, - \, \frac{Q_f^2}{2} e^{\frac{16}{3}f-8w+2\Phi} \, 
-  \, \frac{Q_f^2}{4} e^{\Phi - \frac{4}{3}f-8w} B_{\mu\nu}B^{\mu\nu} \, - \, \frac{Q_c^2}{2} e^{\frac{40}{3}f}
\nonumber \\
&-& 4 Q_f e^{\frac{\Phi}{2} +2f +2w} \sqrt{e^{\Phi + \frac{20}{3}f} + \frac{1}{2} B_{\mu\nu} B^{\mu\nu} } \ , 
\\
{\cal L}_{top} & = & - \, \frac{Q_c}{4} \varepsilon^{\mu\nu\rho\sigma\tau} B_{\mu\nu}\partial_\rho C_{\sigma\tau}  \ ,
\end{eqnarray}
with the convention $\varepsilon^{txyzr}=1$ for the completely antisymmetric symbol. To make contact with the ansatz presented in section \ref{sec:setup} we identify $\kappa_5^2=\kappa_{10}^2/V_{SE}$ and 
\begin{eqnarray}
&& f \, = \,   - \, \frac{1}{5} \log\left[S^4 F h^\frac{5}{4}\right] \,  ,\quad  w \, = \,  \frac{1}{5} \log\left[\frac{F}{S}\right] \ , 
\quad  H_3 \, = \,  \dd B_2 \ , \quad F_3 = \dd C_2   \ ,  
 \\
&&  g_{\mu\nu}\dd x^\mu \dd x^\nu \, \equiv \,  e^{-\frac{10}{3}f}  h^{-\frac{1}{2}} 
\Bigg[ -b_T^2 \dd t^2+b\left( \dd x_1^2+\dd x_2^2 \right) +\dd x_3^2 +  e^{-10f} h^{-\frac{3}{2}} b^2 b_T^2   \dd \sigma^2 \Bigg] \ , 
\nonumber \\
&& B_2=\frac{1}{2}B_{\mu\nu}\dd x^\mu \wedge \dd x^\nu = H(\sigma) \dd x\wedge \dd y \ ,  \quad 
C_2 = \frac{1}{2}C_{\mu\nu} \dd x^\mu \wedge \dd x^\nu = J(\sigma) \dd t \wedge \dd z \ . \nonumber 
\end{eqnarray}
This effective 5d action is not enough to study perturbations, though, since the truncation of fields that cancel in the specific background we are considering is not a consistent one \cite{javialdo}.


\subsection{Perturbative solution} \label{pert-solution}

The system \eqref{defJ} and \eqref{diff-b}--\eqref{diff-H} allows for a systematic expansion of all the functions
in power series of $Q_f$, as defined in equation (\ref{QcQf}). In fact physically it is more relevant to expand in the parameter, $\epsilon_*$
\begin{equation}
{\epsilon_*}\equiv Q_f\, e^{\Phi_*}\ ,\label{epsilonstar}
\end{equation}
which takes into account the running of the effective 't Hooft coupling (through the dilaton factor $e^{\Phi_*}$). 
We consider the following first order expansion in $\epsilon_*$
\begin{eqnarray}
& b \, = \, 1 \, + \, \epsilon_* b_1 +{\cal O}(\epsilon_*^2) \ , \quad 
&h \, = \, \frac{R^4}{r^4} \, \left(1+\epsilon_*h_1 +{\cal O}(\epsilon_*^2)\right) \ , \nonumber \\ 
&S \, = \, r \, \left(1+\epsilon_*S_1 +{\cal O}(\epsilon_*^2)\right)\ ,  \quad
& F\, = \, r \, \left(1+\epsilon_*F_1 +{\cal O}(\epsilon_*^2)\right)  \ , \label{expansion1} \\
&\Phi \, = \, \Phi_*+\epsilon_*\Phi_1 +{\cal O}(\epsilon_*^2) \ ,  \quad 
&H \, = \, H_{0} \left( 1 \, + \, \epsilon_* H_1 +{\cal O}(\epsilon_*^2) \right)   \ .
\nonumber
\end{eqnarray}
where $R^4\equiv Q_c/4$.
We define the new radial coordinate $r$, in such a way that the zeroth order expansion in $\epsilon_* $ of $h$ becomes ${R^4}/{r^4}$
\begin{equation}
e^{-4r_h^4\,\sigma}\, \equiv \, 1 \, - \, \frac{r_h^4}{r^4} \ .
\label{r-def} 
\end{equation}
The extremal limit corresponds to sending the horizon radius $r_h$ to zero.
It is also convenient to define the following parameter
\begin{equation}
r_m^4=e^{-\Phi_*}H_{0}^2R^4 \ ,  \label{dimless-r}
\end{equation} 
The result is a coupled system of second order differential equations which can be decoupled by the transformations
\begin{equation}
\Delta_1 \, \equiv \,  S_1 \, - \, F_1 \ ,  \quad \quad
\Upsilon_1  \, \equiv \, 4\, F_1 \, + \, 16 \, S_1 \, + \, 5 \, h_1 \ , \quad  \quad
\Lambda_1 \, \equiv \, h_1 \, - b_1 \, \ .
\end{equation}
This allows us to write
\begin{eqnarray}\label{eomshot1}
\Psi_1'' \, + \,  \frac{5r^4-r_h^4}{r(r^4-r_h^4)}\Psi_1' \, - \, \frac{4\zeta_\Psi r^2}{r^4-r_h^4} \Psi_1& = & 
\frac{A_\Psi r^4+ B_\Psi r_m^4}{(r^4-r_h^4)\sqrt{r^4+r_m^4}}  \ ,
\\
\tilde H_1'' \, + \, \frac{r^4+3r_h^4}{r(r^4-r_h^4)} \tilde H_1' \, - \, \frac{16 r^2}{r^4-r_h^4}\tilde H_1& = & 
\frac{4r^4
}{(r^4-r_h^4)\sqrt{r^4+r_m^4}}  \ , \label{eomshot2}
\end{eqnarray}
where 
\begin{eqnarray}
\Psi=\{b,\Lambda,\Upsilon,\Delta,\Phi\} \ , & \quad & \zeta_{\{b,\Lambda,\Upsilon,\Delta,\Phi\}}=\{0,8,8,3,0\} \ , 
\nonumber \\
 A_{\{b,\Lambda,\Upsilon,\Delta,\Phi\}}=\{0,0,-16,-1,4\} \ , & \quad & B_{\{b,\Lambda,\Upsilon,\Delta,\Phi\}}=\{-4,2,-6,-1,2\} \ .
\end{eqnarray}
The solution to these equations of motion is described in appendix \ref{blackholeapp}. Let us comment here on the boundary conditions we impose. In our solution there are four scales. We have already introduced the first three: $r_h$ is the radius of the horizon and we impose the fields to be regular  there; $r_m$ is associated to the magnetic field, and $r_*$  denotes the point at which we pierce the dilaton, this is, $\Phi(r)= \Phi_*+\phi(r)$ with $\phi(r_*)=0$. With this scale we defined $\epsilon_*$ and its interpretation is given in terms of the scale at which the gauge coupling is defined, since \cite{Bigazzi:2009bk}
\begin{equation}
\epsilon_* \, = \,  \frac{1}{2 \pi} \, g_s \, N_c \, e^{\Phi_*} \, \frac{N_f}{N_c} \, .
\end{equation}

The fourth scale (which we will define as $r_s$)  is the scale at which we paste the thermal solution presented in the appendix \ref{blackholeapp} to the $T=0$ (supersymmetric) one 
\cite{Bigazzi:2009bk}, \emph{i.e.} we impose the following conditions
\begin{equation}
b_1(r_s)\, = \, H_1(r_s) \, = \, \Lambda_1(r_s) \, = \, 0\ , \qquad  \Upsilon_1(r_s)\, = \, \frac{2}{9} \ ,
 \qquad  \Delta_1(r_s) \, = \, \frac{1}{12} \ . 
\end{equation}

Notice that $b_T(r_s)\neq 1$, which is the supersymmetric solution. This is not a problem in Euclidean signature, since it can be solved by fixing the periodicity of the Euclidean time in the  solution without temperature, and we will use this in the following to compare the energy and free energy of both solutions. In Lorentzian signature it introduces an error of order $(r_h/r_s)^4$, which is small provided $r_h\ll r_s$. 

From now on we set $r_s\to \infty$, which corresponds to push the Landau pole to infinity, or more physically, to focus only in the IR properties of the theory. A UV completion of the system is not known even in the supersymmetric case. The following results can be understood as the leading terms in an $r_h/r_s$ expansion. At the same time, we will take $r_*=r_h$, therefore describing the value of the dilaton relative to its value at the horizon, which implies that the 't Hooft coupling $\lambda_h = 4\pi g_s N_c e^{\Phi_h}$ is evaluated at the energy scale marked by the temperature. 

For completeness, let us mention that from \eqref{defJ} and \eqref{eomshot2} we have at first order in $\epsilon_h$
\begin{equation}\label{Jprimeeom}
\partial_r \tilde J \, = \,  \epsilon_h \left[  \frac{r_m^2 r}{\sqrt{r^4+r_m^4}}-\frac{1}{4}\partial_r \left( \left(1 - \frac{r_h^4}{r^4} \right) r \partial_r \tilde H_1\right) \right]\ .
\end{equation}


\paragraph{Qualitative behavior of the solution}

Given the gargantuan form of the solution to our system at first order in $\epsilon_h$, which can be found in appendix \ref{blackholeapp}, we give in this section a description of the different functions presented above. In this section some numeric work is presented, but in the rest of the paper we will restrict to analytic results.

The function $b_1$ is easy to describe by focusing in its radial derivative,  given by
\be
b_1' = - \frac{2r_m^4}{r\left( r^4-r_h^4\right)} \log \left[ \frac{r^2 + \sqrt{r^4+r_m^4}}{r_h^2 + \sqrt{r_h^4+r_m^4}}  \right] \ ,
\ee
which for $r\geq r_h$ and real non-vanishing $r_m$ is always negative (it is exactly vanishing if $r_m=0$), and asymptotes $b_1'\to0$ at large radius. As the boundary condition used in the integration is  $b_1(r_s)=0$, we conclude that this function is a monotonically decreasing function of $r$ for finite $r_m$ (exactly zero if $r_m=0$) with the maximum value at the horizon.

Similarly, we can analyze the radial gradient of the dilaton correction
\be
\phi_1' = \frac{1}{r} \frac{r^2 r_h^2 + \sqrt{\left( r^4+r_m^4\right) \left( r_h^4+r_m^4\right)}}{r^2\sqrt{r_h^4+r_m^4}+r_h^2 \sqrt{r^4+r_m^4}} \ ,
\ee
which is strictly positive for $r\geq r_h$ and real $r_m$. In this case the boundary condition used to integrate the solution is $\phi_1(r_*)=0$, where $r_*$ will be identified eventually with the horizon position as the IR scale of our effective solution. At large radius the dilaton diverges logarithmically, signaling the presence of a Landau pole, as discussed in \cite{Bigazzi:2009bk}.

\begin{figure}[htb]
\begin{center}
\includegraphics[scale=0.6]{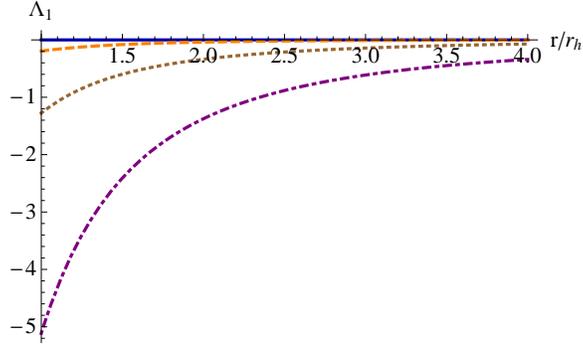}
\caption{$\Lambda_1$ as a function of $r/r_h$ for several values of $r_m/r_h=0$ (blue straight line), $2$ (orange dashed line), $5$ (brown dotted line) and $10$ (purple dotdashed line). To produce this plot  the limit $r_s\to\infty$ has been taken  analytically first.}
\label{fig.lambda1}
\end{center}
\end{figure}

For the other functions present in our solution --namely $\Lambda_1$, $\Upsilon_1$, $\Delta_1$ and $H_1$-- the gradient does not take a simple form that is worth writing, so we provide plots of the functions for several values of the parameters. For example, for $\Lambda_1$ one has that, numerically, the radial gradient is strictly non-negative (zero if $r_m=0$), and $\Lambda_1(r_s)=0$ from the boundary condition, in a similar situation to the function $b_1$ but with different sign for the gradient. In figure \ref{fig.lambda1} we plot  this quantity as a function of $r/r_h$ for several values of $r_m/r_h=0,2,5,10$ and observe that it has non-negative gradient, and approaches $\Lambda_1\to 0$ as $r\to r_s$ (with $r_s\to\infty$ in the figure).

\begin{figure}[htb]
\begin{center}
\includegraphics[scale=0.6]{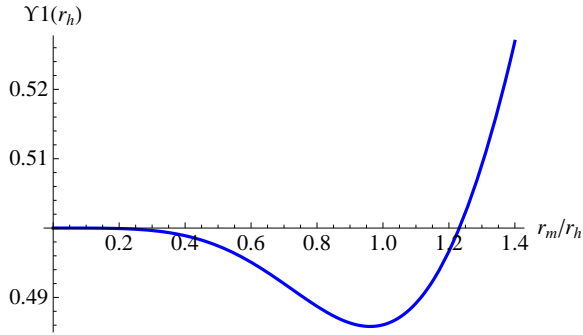}
\caption{$\Upsilon_1(r_h)$ as a function of $r_m/r_h$. We observe a minimum at $r_m=0.961122 r_h$ with value $\Upsilon_1=0.485816$ and the curve crosses $\Upsilon_1=\frac{1}{2}$ again at $r_m=1.23144 r_h$. To produce this plot  the limit $r_s\to\infty$ has been taken  analytically first.}
\label{fig.horupsilon}
\end{center}
\end{figure}

As opposed to the previously presented cases, function $\Upsilon_1$ presents some  structure. To start with, the boundary condition at $r=r_s$ changes and is given by $\Upsilon(r_s)=2/9$. However, when one works in the $r_s\to \infty$ limit this boundary condition is modified to $\Upsilon_1(\infty)=1/2$, which is the value of the function when $r_m=0$. For small values of the magnetic field scale (weighted by the horizon radius), $r_m/r_h\lesssim 1.23144$, the value of $\Upsilon_1$ at the horizon is less than $1/2$, and after that specific value of the magnetic scale it is always larger than $1/2$. We plot this behavior in figure \ref{fig.horupsilon}. Given the analyticity of the function there is a minimum which, numerically, we determined to be at $r_m\approx 0.961122 r_h$.
We have not found any characteristic signature of the presence of this minimum of $\Upsilon_1(r_h)$ in the plasma.

\begin{figure}[htb]
\begin{center}
\subfigure[$0\leq r_m\leq 0.961122 r_h$]{\includegraphics[scale=0.6]{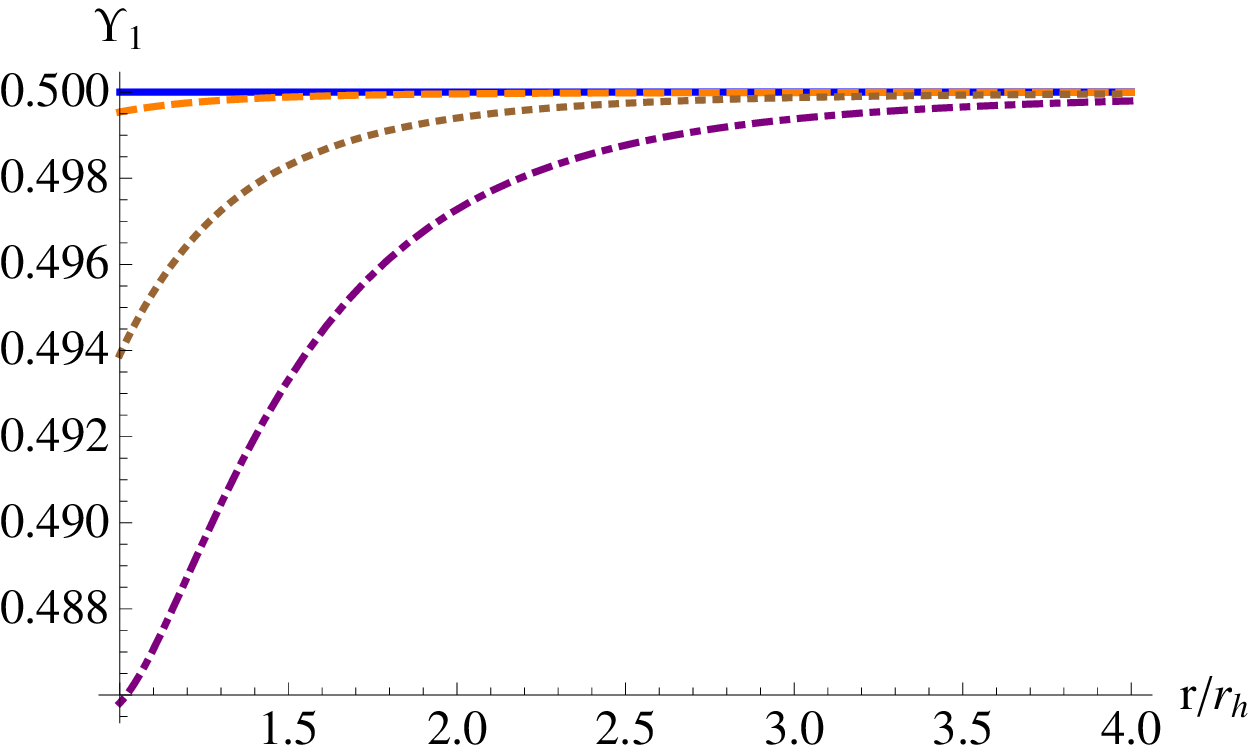}}
\subfigure[$0.961122 r_h\leq r_m \leq 1.23144r_h$]{\includegraphics[scale=0.6]{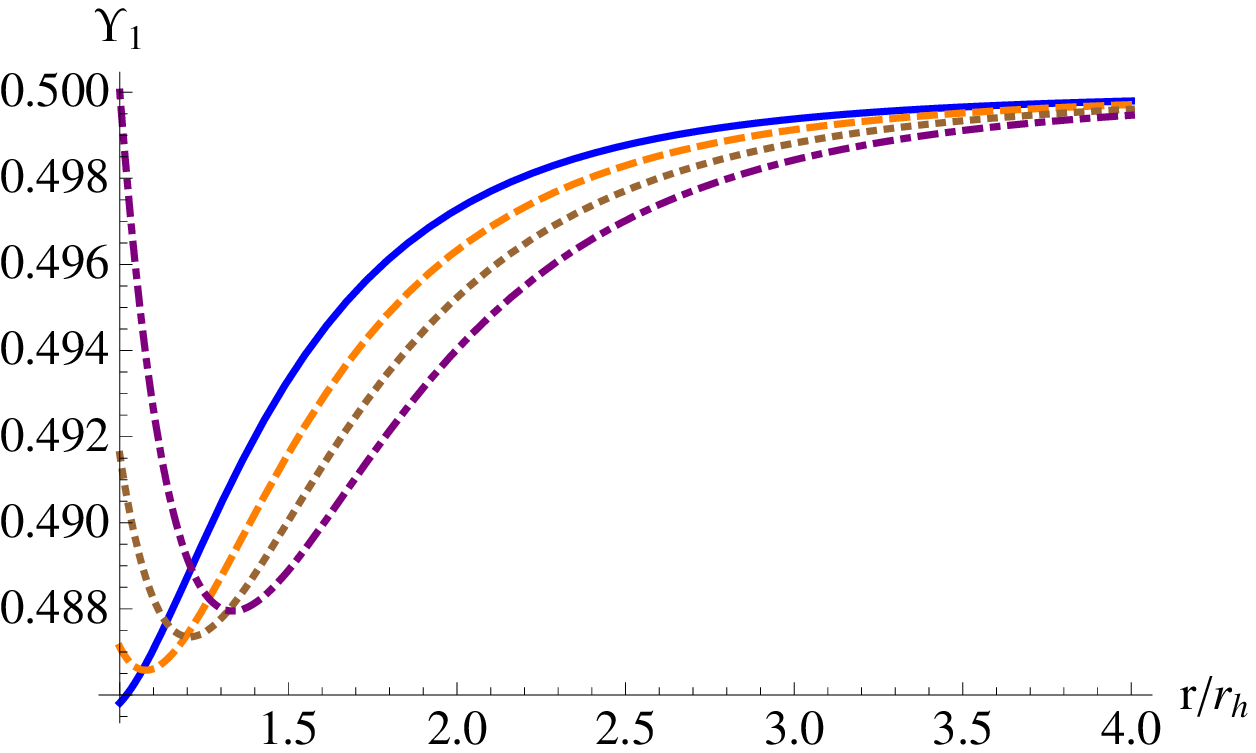}}
\subfigure[$r_m \geq 1.23144 r_h $]{\includegraphics[scale=0.6]{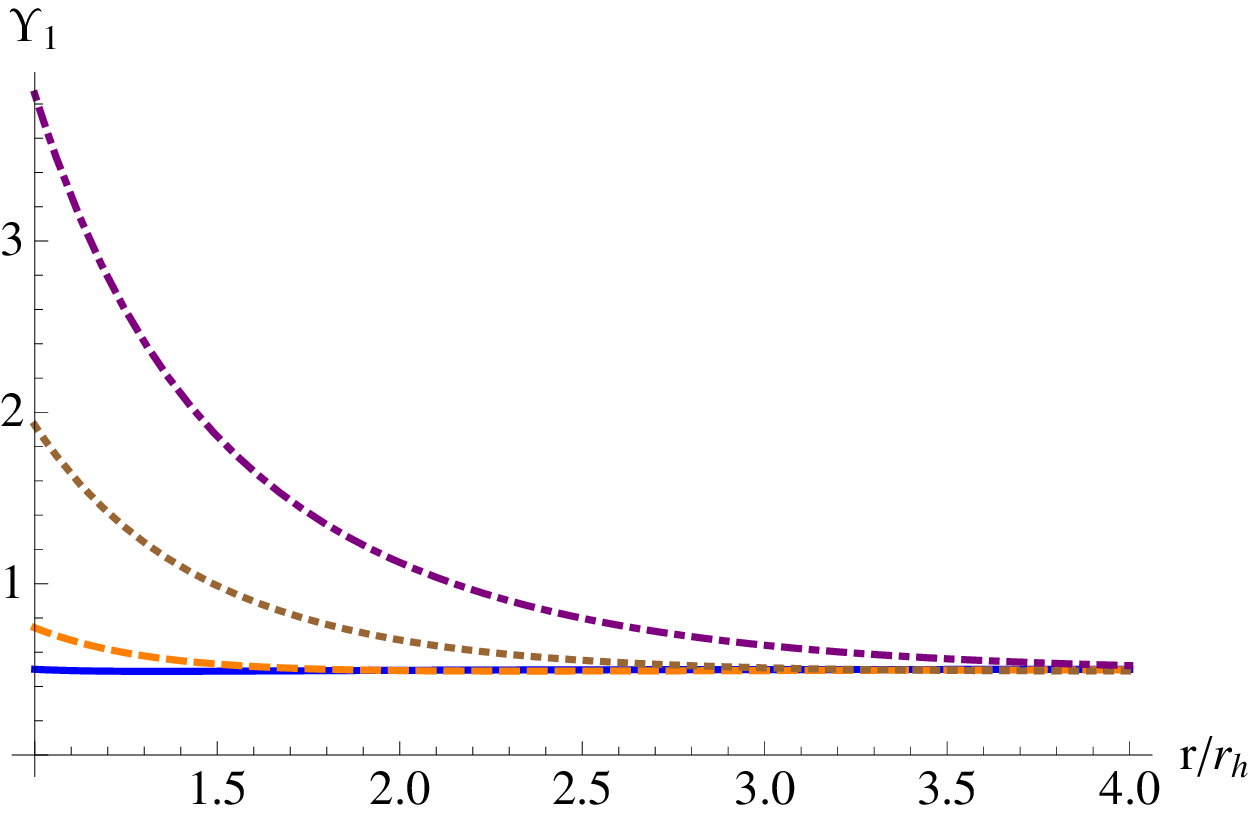}}
\caption{$\Upsilon_1$ as a function of $r/r_h$ for several values of $r_m/r_h$ represented by a blue straight line,  an orange dashed line, a brown dotted line, and a purple dotdashed line, with values given respectively by (a) $0,\, 0.32,\, 0.64,\, 0.961122$, (b) $0.961122, \, 1.05, \,1.14, \, 1.23144$ and (c) $1.23144, \, 2, \, 3.5, \, 5$. To produce this plot the limit $r_s\to\infty$ has been taken analytically first.}
\label{fig.upsilon1}
\end{center}
\end{figure}

In figure \ref{fig.upsilon1} we plot several examples of $\Upsilon_1$ as a function of the radial variable in three graphs, classified according to the value of the function at the horizon. All the curves present a minimum (on the horizon when $r_m\leq 0.961122 r_h$ and on the bulk otherwise) and asymptote the $r_m=0$ value ($\Upsilon_1=1/2$) at large radius. 

\begin{figure}[htb]
\begin{center}
\includegraphics[scale=0.6]{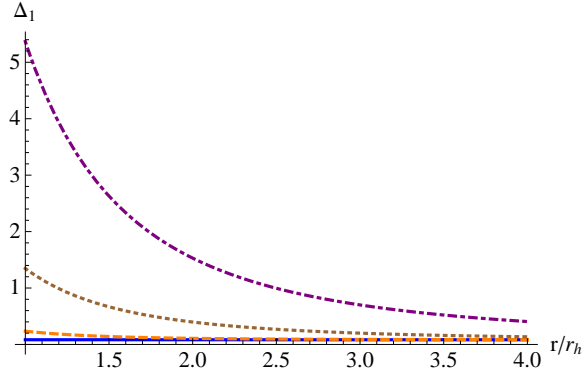}
\caption{$\Delta_1$ as a function of $r/r_h$ for several values of $r_m/r_h=0$ (blue straight line), $2$ (orange dashed line), $5$ (brown dotted line) and $10$ (purple dotdashed line). To produce this plot $r_s$ has been taken to $r_s=80 r_h$ in the numerics. We have checked that this value for $r_s$ gives indistinguishable results from those in figures \ref{fig.lambda1} and \ref{fig.upsilon1}.}
\label{fig.delta1}
\end{center}
\end{figure}

We have not given an analytic expression for $\Delta_1$ because we couldn't find an easy way to write it, since it involves integrals of Legendre functions. However, from integrating the equation numerically we find that its behavior is very similar to that of $b_1$ or $\Lambda_1$ (with reversed sign), and we simply report here figure \ref{fig.delta1}. We are not going to need to evaluate $\Delta_1$ anywhere in this work. The reason is that this is the mode describing the squashing in the compact Sasaki-Einstein manifold, but from the point of view of the 5-dimensional system it is just a scalar that does not enter explicitly in the 5-dimensional metric (its influence would be felt just via the equations of motion, but recall  we have defined $\Delta_1$ precisely to decouple them). In this paper we will focus on the  thermodynamics, stress-energy tensor and energy-loss of probes in the system, which do not depend explicitly in the matter content of our theory, just in the 5-dimensional metric.

\begin{figure}[htb]
\begin{center}
\includegraphics[scale=0.6]{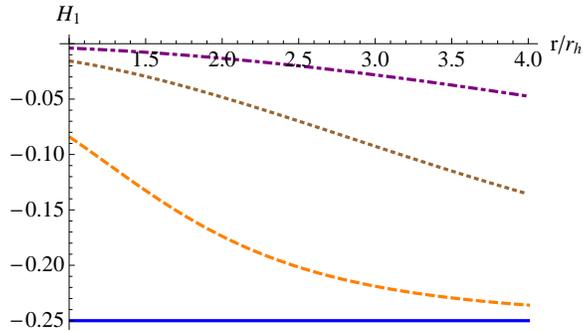}
\caption{$H_1$ as a function of $r/r_h$ for several values of $r_m/r_h=0$ (blue straight line), $2$ (orange dashed line), $5$ (brown dotted line) and $10$ (purple dotdashed line). To produce this plot  the limit $r_s\to\infty$ has been taken  analytically first.}
\label{fig.h1}
\end{center}
\end{figure}

Finally, we present the flavor correction to the NSNS 2-form $H_1$. As usual we take the $r_s\to\infty$ limit analytically and we find, as it was the case for $\Upsilon_1$, that the boundary condition is not $H_1(\infty)=0$ but $H_1(\infty)=-1/4$. One might be puzzled by the fact that the correction is not vanishing independently of the value of $r_m$, in concrete when $r_m=0$, but recall that this correction is modulated by the flavorless value of the NSNS field strength $H\sim r_m^2 (1+\epsilon_h H_1+{\cal O}(\epsilon_h)^2)$, therefore at vanishing magnetic field we have $H=0$. As the value of the magnetic field is increased the correction gets smaller and smaller as can be seen in figure \ref{fig.h1}.


\subsection{Hierarchy of scales and regime of validity of the supergravity solution}\label{sec.regime}

The perturbative solution, that we present in full detail in the appendix \ref{blackholeapp}, needs to be supplemented with a hierarchy 
of energy scales  (in terms of radial scales). Our analysis follows closely similar sections of \cite{Bigazzi:2009bk,Filev:2011mt,Erdmenger:2011bw}, whose arguments we repeat here for completeness.

As usual, for the Taylor expansions in \eqref{expansion1} to be valid in the region $r_h  \leq r\leq r_s$ 
we need to separate the scale $r_s$ from the scale introduced by the solution to $\phi_1(r)$, which diverges logarithmically at large values of the radius $r> r_s$,
$
r_s\ll r_h e^{1/\epsilon_h}  
$. 
 The requirement that we discard corrections in $r_h/r_s$ implies that our perturbative corrections are much larger than the terms we discard, therefore
$
\epsilon_h \gg {r_h}/{r_s}
$. 
 Joining these two conditions we have
\begin{equation}
e^{-1/\epsilon_h}  \ll  \frac{r_h}{r_s} \ll \epsilon_h \ ,
\end{equation}
which for large $r_s\gg r_h$ (implying that the UV completion this theory needs is far from the IR, where we study the physical properties of the system) implies that
\begin{equation}
0< \epsilon_h \sim \lambda_h \frac{N_f}{N_c} \ll1 \ .
\end{equation}

The scale $r_m$ is associated with the magnetic field and can be arbitrarily close to $r_m=0$. For large values of $r_m$ (large magnetic fields/magnetization of the system, as we will see in the next section), requiring that our solution remains in the perturbative level sets up a top value. As can be seen from the plots given previously, the maximum value of the functions appearing in the solution is at the horizon, and from the asymptotic values given in appendix \ref{blackholeapp} it is easy to see that, at large values of $r_m$, all the functions diverge at the horizon as $r_m^2/r_h^2$. Therefore we must impose $1\gg \epsilon_h r_m^2/r_h^2$, which gives the condition
\begin{equation}
|r_m| < \frac{r_h}{\epsilon_h^{1/2}} \ .
\end{equation}

Similarly to \cite{Bigazzi:2009bk}, validity of the supergravity approximation requires to ignore closed string loops ($N_c\gg 1$) and 
$\alpha'$ corrections ($\lambda_h \gg 1$), where $\lambda_h$ is the effective 't Hooft coupling at the energy scale set by the temperature. 
In addition, validity of the smearing approximation suggests a dense distribution of flavor D7-branes. 
In summary we have
\begin{equation} 
\{N_c, \, N_f\} \gg 1,\quad \lambda_{h}\gg 1\ ,\quad \epsilon_{h}\equiv \frac{\lambda_h\,N_f}{8\pi^2\,N_c}\ll 1\ .
\end{equation}
Finally requiring that $\alpha'$ corrections, which scale as $\lambda_h^{-3/2}$, 
are sub-leading relative to flavor corrections, controlled by $\epsilon_h$, requires
\begin{equation}
\lambda_h^{-3/2}\ll \epsilon_h \ .
\end{equation}


\section{Thermodynamics}\label{sec:TD}

In the previous section we presented in full detail the solution of an anisotropic black hole and now we 
will extract its thermodynamic properties. This will provide a non trivial validity check of the solution itself through the 
closure of the standard thermodynamical formulae. As in \cite{Bigazzi:2009bk, Bigazzi:2011it}, all quantities are obtained in power series of the 
perturbative expansion parameter and, therefore, the relevant  thermodynamic relations are verified up to the relevant order.


\subsection{Smarr formula}

The temperature of the black hole is computed after imposing regularity of the Euclidean action. 
A simple computation using  \eqref{limit-b1},  \eqref{limit-L1} and \eqref{limit-Y1} gives\footnote{In this section all quantities have corrections coming from $ {\cal O}(\epsilon_h^2)$ terms as well as $\frac{r_h^4}{r_s^4}$, where we are setting $r_s\to\infty$.}
\begin{equation}
T  =  \frac{r_h}{\pi R^2} \Bigg[ 1 \,  + \,  \frac{1}{4} \, 
\epsilon_h \,  \left(\,  3 \Lambda_1 \, -  \, \Upsilon_1 \, - \,  b_1 \right) \Bigg]_{r = r_h} \!\!\!\! =  \frac{r_h}{\pi R^2} \left[ 1 \,  + \,  \frac{1}{8} \, 
\epsilon_h \,  \left(\,  1 \, -  2 \, \sqrt{1\, + \, \frac{r_m^4}{r_h^4}} \right) \right] \, .  \label{temperature}
\end{equation}
The entropy density is proportional to $A_8$, the volume at the horizon of the eight dimensional part of the space orthogonal to the $\hat t,r$ 
plane (where $\hat t$ is the Euclidean time), divided by the infinite constant volume of the 3d space directions
$V_3$. Another simple computation using  \eqref{limit-b1},  \eqref{limit-L1} and \eqref{limit-Y1} gives
\begin{eqnarray}
s & = & \frac{2\pi}{\kappa_{10}^2} \,  \frac{A_8}{V_3} \, = \, 
\frac{N_c^2 \, r_h^3}{2 \pi  \, R^6} 
 \Bigg[ 1 \,  - \,  \frac{\epsilon_h }{4} 
 \,  \left(\,  3 \Lambda_1 \, -  \, \Upsilon_1 \, - \,  b_1 \right) \Bigg]_{r = r_h}\!\!\!\! = \frac{N_c^2 \, r_h^3}{2 \pi  \, R^6} \left[ 1 \,  - \,  \frac{\epsilon_h }{8} 
 \,  \left(\,  1 \, -  2 \, \sqrt{1\, + \, \frac{r_m^4}{r_h^4}} \right) \right]  \nonumber \\ \nonumber \\ 
&=&   \frac{N_c^2 \pi^2 T^3}{2} \Bigg[ 1 \,  + \,  \frac{\epsilon_h }{2}\,  \left(\,  1 \, -  2 \, \sqrt{1\, + \, \frac{r_m^4}{r_h^4}} \right)  \Bigg]\ .  \label{entropy1}
\end{eqnarray}
Note also combining \eqref{temperature} and \eqref{entropy1} that 
\begin{equation}\label{tempentropy}
s \, T \, = \, \frac{2 \, \pi^3}{\kappa_{10}^2} r_h^4 \ .
\end{equation}
In principle this result is perturbative in $\epsilon_h$ and valid to order $\epsilon_h^2$ in the present case, however, it is not difficult to show that the statement is true, independently of the expansion parameter.

We define now the magnetic quantities. One natural identification for the magnetic field, $B$, is given by the value of the $H$ field at the boundary, which from \eqref{dimless-r} is\footnote{Notice that we cancel a factor of $e^{\Phi_h}$ by passing between the string and Einstein frames.}
\begin{equation} \label{defB}
B \, = r_m^2 R^{-2} \,  .
\end{equation}
Looking at \eqref{defJ}, we see that $J'$ and $H_{0}$ are conjugate variables. The existence of the holographic duality implies that, 
if we associate $H_{0}$ with the magnetic field then $J$ has to determine the magnetization density $\cal M$. This relation  reads
\begin{equation}  \label{magnetization1}
{\cal M} \, \equiv \, \frac{1}{V_3} \int \frac{\delta  S_{eff}}{\delta H_{0}} \,= \, - \frac{Q_c \, \pi^3}{2 \kappa_{10}^2} \int J' dr 
 \,= \, - \frac{N_c^2}{2 \pi^2 R^4}  \, \Delta J _{reg} \ , 
\end{equation}
and we will regularize the finite temperature result subtracting the zero temperature one. Using  \eqref{Jprimeeom}, to obtain the integral of $J'$,
we arrive to the following expression for the magnetization 
\begin{equation}  \label{magnetization2}
{\cal M} \, = \, \frac{N_c^2}{2 \pi^2 R^4} \, \frac{Q_f \, B}{2} \, \log \left[ \frac{r_h^2+\sqrt{r_h^4+r_m^4}}{r_m^2} \right] \,  .
\end{equation}
The next step in the determination of the Smarr formula is the calculation of the internal energy. Starting from the ADM energy we have 
\begin{equation}\label{ADMdef}
E_{ADM} \,  =  \, - \, \frac{1}{\kappa_{10}^2} \sqrt{-g_{tt}} \int d^8x \sqrt{\det g_8} (K_T-K_0)\ .
\end{equation}
The eight-dimensional integral is taken over a constant time, constant radius hypersurface.
The symbols $K_T$ and $K_0$ are the extrinsic curvatures of the eight-dimensional
subspace within the nine-dimensional (constant time) space, at finite  and
zero temperature, respectively. Using the explicit solution in appendix \ref{eff-EOM} we have 
\begin{equation} \label{ADM}
\varepsilon_{ADM} \, = \, 	\frac{E_{ADM}}{V_3} \, = \,  \frac{3 \, N_c^2 \, r_h^4}{8 \, \pi^2 \, R^8} \, 
\left[ 1+ \epsilon_h \,  \frac{r_m^4}{3 \, r_h^4} \log \left[ \frac{r_m^2}{r_h^2+\sqrt{r_h^4+r_m^4}} \right]  \right] \ . 
\end{equation}
Another way to write \eqref{ADM} is, at order ${\cal O}(\epsilon_h^2)$,
\begin{equation} \label{ADM2}
\varepsilon_{ADM} \, = \, \frac{3}{4} \, s \, T \, - \,  \frac{1}{2}\, B \, {\cal M} \ , 
\end{equation}
which implies that we must identify the ADM mass with the magnetic enthalpy of the system, ${\cal H}=\varepsilon_{ADM}$. 
The internal energy, $\cal U$, is given in terms of the enthalpy by the following expression 
\begin{equation}\label{internalenergy}
{\cal U} \, = \,  {\cal H} \, + \,  B \,  {\cal M} \, = \,  \frac{3}{4}\,  s \, T \, + \, \frac{1}{2} \,  B \, {\cal M} \ .
\end{equation}
%


\subsection{Thermodynamic potentials}

The  relations that must be satisfied by the thermodynamic potentials are the following
\begin{equation}
{\cal F} \, = \, {\cal U} \, - \,  s \, T \  , \quad  \quad  {\cal G} \,  = \,  {\cal F} \, - \,  B \, {\cal M} \ ,
\end{equation}
where $\cal F$ is the Helmholtz free energy (in the ensemble where the magnetization is kept fixed) and 
$\cal G$ is the Gibbs free energy (in the ensemble where the magnetic field is kept fixed), which is the interesting ensemble in our case. 
These thermodynamic potentials are related by a Legendre transformation
\begin{equation}
{\cal G} \, = \, {\cal F} \, - \,  \frac{\partial {\cal F}}{{\partial \cal M}} \, {\cal M} \ .
\end{equation}
In a holographic set-up the thermodynamic potentials are related to the on-shell Euclidean action (times the temperature to cancel the periodicity of the Euclidean time direction). 
Given our previous discussion on the identification of the magnetization with the field $J$ (see \eqref{magnetization1}), which leads to the following relation 
\begin{equation}
\frac{\partial {\cal M}}{\partial J}  \, = \,  \frac{{\cal M}}{J} \ ,
\end{equation}
we can associate the Legendre transformation defining the Gibbs free energy with the Legendre transformation defining the action $\tilde S$ in \eqref{L-effective2}. 
By denoting the on-shell action\footnote{Notice that to obtain $\tilde S_{eff}$ we have integrated by parts to get rid of second order differentials, introducing some boundary terms. 
These, in principle, are taken care of by the Gibbons-Hawking term and will not contribute to the final expression. 
We have checked that this is the case by calculating the on-shell action with and without these extra boundary terms.} 
as $\tilde I$ we have\footnote{This calculation is detailed in appendix \ref{Gpotential}.}
\begin{equation} \label{gibbsenergy}
{\cal G} = \frac{\tilde I}{\beta} = 
- \, \frac{1}{8} \, N_c^2 \, \pi^2 \,  T^4  \left[ 1 \, + \, \epsilon_h \, \left( - \, \frac{1}{2} + \sqrt{1+\frac{r_m^4}{r_h^4}} + \frac{r_m^4}{r_h^4}\, 
\log \left[ \frac{r_h^2+\sqrt{r_h^4+r_m^4}}{r_m^2} \right] \right) + {\cal O}(\epsilon_h^2)  \right] \ . 
\end{equation}
The regularization is performed by subtracting the $T=0$ background and the action is supplemented with a Gibbons-Hawking term. 
In \cite{Erdmenger:2011bw,Albash:2007bk} the regularization of the free energy in the probe approximation was performed by the addition of counterterms. We consider the fact that we recover their results in the appropriate limit as a sign that the background subtraction method gives the correct answer. In particular, we do not need to worry about the presence of logarithmic divergences, cancelled by counterterms with explicit cutoff dependence, since these are temperature independent and the background subtraction cancels them completely.
It is not difficult to check that, when subtracting the $T=0$ background, the contribution of that term vanishes up to order $1/r_*^4$. 
Now, it is not difficult to check that indeed
\begin{equation}
{\cal G}  \, = \, - \, \frac{1}{4} \, s \, T \, - \, \frac{1}{2} \, { B} \,  {\cal M} \ .
\end{equation}
Applying the standard thermodynamic relations
\begin{equation}
s \, =  \, - \,  \left(\frac{\partial {\cal G}}{\partial T}\right)_{B} \ , \quad  \qquad  
{\cal M} \, =  \, - \,  \left(\frac{\partial {\cal G}}{\partial {B}}\right)_{T}  \ , 
\end{equation}
we confirm the previously obtained results in \eqref{entropy1} and \eqref{magnetization2}  respectively.
The calculation of the Helmholtz free energy can be done in a similar fashion, but  using the original action \eqref{eff-action}.
This can be seen as the Legendre transformation of $\tilde S_{eff}$, which eliminates $H_{0}$ and adds a term $+ \, {B}{\cal M}$, after the 
proper renormalization.  In this way we have 
\begin{equation} \label{helmholtzenergy}
{\cal F} \,  =  \, {\cal G}  \, + \,  {B}{\cal M} \, = \, {\cal U} \, -  \, s \, T  \quad \Rightarrow  \quad 
{\cal F}  \, =  \, - \, \frac{1}{4} \, s \, T \, + \,  \frac{1}{2} \, {B} \, {\cal M} \ .
\end{equation}
Once again we can check the thermodynamic relations
\begin{equation}
s \, =  \, - \,  \left(\frac{\partial {\cal F}}{\partial T}\right)_{\cal M} \ , \quad  \qquad  
B \, =   \,  \left(\frac{\partial {\cal F}}{\partial {\cal M}}\right)_{T}  \ , 
\end{equation}
where, to work at fixed magnetization, we have to specify how $r_m$ evolves with the temperature. For that we look at equation \eqref{magnetization2}, from where the following evolution follows
\begin{equation}\label{magnetscal}
\partial_T\, r_m = \frac{\pi R^2 r_h r_m}{r_h^2-\sqrt{r_h^4+r_m^4} \log \left[ \frac{r_h^2+ \sqrt{r_m^4+r_h^4}}{r_m^2} \right]} \ .
\end{equation}
%


\subsection{Speed of sound}

Finally we analyze the speed of sound in the plasma with a magnetic field. Due to the anisotropy of the gravitational solution we will find  that there are two normal directions in which the pressure waves propagate at different speeds. For a perturbation propagating in the direction of the magnetic field we have
\begin{equation}
c_{s,||}^2 = \frac{\partial P_{||}}{\partial {\cal U}} = \frac{- \left( \partial {\cal G} / \partial T \right)_{B}}{\left( \partial {\cal U} / \partial T \right)_{B}} = \frac{s}{C_{V,B}} \ ,
\end{equation}
where $C_{V,B}$ is the heat capacity at fixed magnetic field. To calculate it we have to derive the internal energy with respect to the temperature, but we must take into account how the parameters $\epsilon_h$ and $r_m$ run with the energy scale. The case of the parameter $\epsilon_h$ is easy to understand from the profile for the dilaton and it follows that 
$\partial_T \epsilon_h \sim \epsilon_h^2$, \cite{Bigazzi:2009bk}. Since we work at first order in $\epsilon_h$, the running of the coupling constant\footnote{Recall that 
$\epsilon_h \sim \lambda_h N_f/N_c$.} -- via the presence of factors of $R$ in the definitions of the physical magnetic field  \eqref{defB} and the magnetization \eqref{magnetization2} --  does not affect our results. At fixed magnetic field, since $B\sim r_m^2$ we observe that $\partial_T \, r_m=0$, therefore
\begin{equation}
C_{V,B} = \left( \frac{\partial {\cal U}}{\partial T} \right)_B = \frac{3 N_c^2 r_h^3}{2 \pi R^6} \left[ 1 - \frac{\epsilon_h}{8} \left( 1- \left( 2 +  \frac{10}{3} \frac{r_m^4}{r_h^4} \right) \frac{1}{\sqrt{1+\frac{r_m^4}{r_h^4}}}   \right) + {\cal O}(\epsilon_h^2) \right] \ .
\end{equation}
With this result at hand we can find readily the speed of sound in the direction of the magnetic field as
\begin{equation}\label{cspara}
c_{s,||}^2 = \frac{s}{C_{V,B}} = \frac{1}{3} \left[  1 - \frac{\epsilon_h}{6} \frac{r_m^4}{r_h^4}  \frac{1}{\sqrt{1+\frac{r_m^4}{r_h^4}}}   + {\cal O}(\epsilon_h^2) \right] \ ,
\end{equation}
which gives a lower speed of sound than the conformal result. For the speed of sound in the direction orthogonal to the magnetic field we 
obtain, using the chain rule
\begin{equation}
c_{s,\perp}^2 =  \frac{\partial P_{\perp}}{\partial {\cal U}} = \frac{- \left( \partial {\cal F} / \partial T \right)_{B}}{\left( \partial {\cal U} / \partial T \right)_{B}} =  -\frac{\left( \frac{\partial {\cal F} }{ \partial T} \right)_{\cal M} + \left( \frac{\partial {\cal F} }{ \partial {\cal M}} \right)_{T} \left( \frac{\partial {\cal M} }{ \partial T} \right)_{B}}{\left( \partial {\cal U} / \partial T \right)_{B}} = \frac{s}{C_{V,B}} - \frac{B}{C_{V,B}} \left( \frac{\partial {\cal M} }{ \partial T} \right)_{B} \ ,
\end{equation}
which leads to
\begin{equation}\label{csortho}
c_{s,\perp}^2 = \frac{1}{3} \left[  1 - \frac{7\,\epsilon_h}{6} \frac{r_m^4}{r_h^4}  \frac{1}{\sqrt{1+\frac{r_m^4}{r_h^4}}}   + {\cal O}(\epsilon_h^2) \right] \ .
\end{equation}
 In particular we see that the presence of a magnetic field in our setup breaks conformal invariance at first order in $\lambda_h \frac{N_f}{N_c}$ even when the fundamental degrees of freedom we included are massless (in the absence of magnetic field the breaking of conformal invariance happens at order $\epsilon_h^2$, see 
\cite{Mateos:2007vn, Bigazzi:2009bk,Bigazzi:2011it,hydro}. This is one difference between the setup presented in this work and the results in the quenched approximation $\lambda_h N_f/N_c\to0$, \cite{Albash:2007bk}. Although for thermodynamic quantities such as the entropy, the magnetization and the Gibbs free energy we obtain agreement with the results of that paper, in our setup the anisotropy sourced by the magnetic field is included, and this allows us to calculate the different speeds of sound, depending on the direction of the pressure wave, and obtain conformality-breaking results.

For completeness we calculate here the heat capacity at constant magnetization, where we need to make use of equation \eqref{magnetscal} to work at fixed magnetization. The result is
\begin{eqnarray}
C_{V,{\cal M}} & = &  \left( \frac{\partial {\cal U}}{\partial T} \right)_{\cal M} = \frac{3 N_c^2 r_h^3}{2 \pi R^6} \Bigg[ 1 +  \frac{\epsilon_h}{24} \, C_{V,{\cal M}}^{cor}
+ {\cal O}(\epsilon_h^2) \Bigg] \ ,
\end{eqnarray}
with
\begin{eqnarray}
C_{V,{\cal M}}^{cor} &= & \frac{1}{{r_h^4 \sqrt{1+\frac{r_m^4}{r_h^4}} \left( r_h^2-\sqrt{r_h^4+r_m^4} \log \left[ \frac{r_h^2+ \sqrt{r_m^4+r_h^4}}{r_m^2} \right] \right)} } 
\, \left[  6r_h^2 \left(r_h^4+3 r_m^4 \right) -3 r_h^4 \sqrt{r_h^4+r_m^4}  \right.
\nonumber  \\
&+& \left. \left( 3r_h^2 \left(r_h^4+r_m^4 \right) -2 \left(3 r_h^4+r_m^4\right) \sqrt{r_h^4+r_m^4} \right) \log \left[ \frac{r_h^2+ \sqrt{r_m^4+r_h^4}}{r_m^2} \right]  \right] \ .
\end{eqnarray}
%


\section{Stress-energy tensor with a magnetic field}\label{sec:setensor}

In this section we will calculate holographically the stress-energy (SE) tensor of the boundary field theory. As customary in the AdS/CFT context, we evaluate the Brown-York tensor at a cutoff $r_\Lambda$ from the 5d action \eqref{5daction}
\be\label{bytensor}
 \tau^{ij}  = \frac{2}{\sqrt{-\gamma}} \frac{\delta S_{5d}}{\delta \gamma_{ij}} \Bigg|_{r_\Lambda} = \frac{1}{\kappa_5^2} \left( K^{ij}-K \gamma^{ij} \right)_{r_\Lambda}\ ,
\ee
where $\gamma$ is the induced metric at the $r=r_\Lambda$ surface, where the indices $i, j$ run, and $K^{ij}$ is the extrinsic curvature. The Brown-York tensor diverges when the cutoff is taken to the boundary. To cancel this divergence we employ the same background subtraction as in the previous section, which allows us to read the temperature and magnetic field contribution to the SE tensor.

\subsection{Expectations}

Before presenting the actual calculation we will state what we \emph{expect}  the diagonal components of the SE tensor to be. The presence of the magnetic field sources an anisotropy in the medium, and therefore we will have a vev for the SE tensor of the field theory given by
\begin{equation}\label{setensor}
\langle T^i{_j} \rangle = \mathrm{diag}\left( - {\cal E}_{ADM}, P_\perp, P_\perp, P_{||} \right) \ ,
\end{equation}
where ${\cal E}_{ADM}$ is the enthalpy, as was shown in the previous section, $P_\perp$ the pressure in the directions transverse to the magnetic field, and $P_{||}$ the pressure along the direction of the magnetic field.  When no magnetic field is present the two pressures coincide and are related to the Gibbs free energy\footnote{In the absence of a magnetic field the Gibbs and Helmholtz free energies presented in \eqref{gibbsenergy} and \eqref{helmholtzenergy} coincide, but the presence of non-trivial charge density and chemical potential would make a difference between the two. Actually, in a traditional nomenclature the Gibbs free energy should correspond to the thermodynamic potential at fixed chemical potential and zero magnetic field; we use the same name here by analogy.} $P_\perp=P_{||}=-{\cal G}$.

The question that is immediately risen is whether in our case $P_\perp=-{\cal G}$ or $P_{||}=-{\cal G}$ --if any--, and if this is true what is the expression for the other pressure. Notice that the difference $\Delta_P\equiv P_{||}-P_\perp$ is a measure of the anisotropy of the medium, and therefore we expect it to be proportional to the magnetization (times the magnetic field) $\Delta_P \sim B{\cal M}$.

The answer to this question is given by
\begin{equation}\label{thermopots}
P_{||}  = -{\cal G}  \ , \qquad P_\perp = - {\cal F}  \ .
\end{equation}
To understand why this is the case, we will follow  a thermodynamic argument that can be found in a similar context in appendix C of \cite{Mateos:2011tv}. In that paper the thermal ${\cal N}=4$ SYM plasma has an anisotropy sourced by a specific distribution of D7 branes along the horizon of the black brane, translated in a value for the axion $\chi=a z$, with $a$ a constant. In the present case the D7 branes are extended along the radial direction of AdS, reaching the boundary and describing fundamental matter in the plasma, and the anisotropy appears by the presence of a magnetization of the fundamental.

The key of the argument is to write the internal energy of the plasma as an \emph{extensive} quantity\footnote{Note that in the rest of the paper thermodynamic quantities are intensive!}
$
{\cal U} = {\cal U}(S, L_x, L_y, L_z,{\cal M})
$, with $S$ the extensive entropy, $L_{x,y,z}$ the length of the sides of a box in which we have inserted our plasma and $\cal M$ the magnetization of the system. The energy and the entropy scale with the total volume of the box $V_3=L_x L_y L_z=\int d^3 x$, but the magnetization does not. This may seem strange at first sight, since one would expect the magnetic field to be an intrinsic quantity and the magnetization to be a density. One way to see the scaling is to realize that the magnetization is a vector in the $z$ direction whereas the magnetic field is given by a $2$-form ${B}\, dx \wedge dy$. Therefore, to keep the magnetic field constant when we scale $L_x$ or $L_y$, we should scale $B$ accordingly. In the same way, the magnetization $\cal $ scales with $L_z$. This suggests that it is more appropriate to talk about magnetic flux along the xy plane, {B}, and magnetization linear density along the $z$ direction. These are the quantities that matter when considering a finite box is the presence of an external magnetic field.

Therefore, comparing with the calculation in \cite{Mateos:2011tv}, all we need to do is to repeat the arguments in their appendix C with the identification $a \to {\cal M}$ and $\Phi \to {B}$ --which we will not write explicitly here since it is nicely discussed in the referred paper--, and we are led to the result \eqref{thermopots}. From here, it is also straightforward to see that $\Delta_P={B}{\cal M}$. 

The identities (\ref{thermopots})  can also be written as Gibbs-Duhem equations
\begin{equation}
{\cal U} + P_{||} = sT + {B}{\cal M} \ , \qquad {\cal U} + P_\perp = sT \ .
\end{equation}


\subsection{Holographic calculation}\label{stressenergy}

We proceed now to calculate the components of the vev of the SE tensor in the field theory. This is related to the Brown-York tensor \eqref{bytensor} by
\be
\langle T^i{_j} \rangle =  \sqrt{-\gamma} \,\tau^i{_{j,reg}}\Big|_{r_\Lambda \to \infty}  \ ,
\ee
where we have assumed that the expression \eqref{bytensor} has been regularized before taking the $r_\Lambda \to \infty$ limit. Notice that strictly speaking this is a density since we are not integrating over the space. From the definition \eqref{bytensor} we have
\beqa
\langle T^t{_t} \rangle  & = & \frac{1}{2\kappa_5^2} r^5 \left( 1-\frac{r_h^4}{r^4} \right)  \partial_r \log \left(e^{-10f}h^{-3/2} b^{2}\right)    \ ,  \label{bytt} \\
\langle T^x{_x} \rangle =  \langle T^y{_y} \rangle & = & \frac{1}{2\kappa_5^2} r^5 \left( 1-\frac{r_h^4}{r^4} \right) \partial_r \log 
\left( e^{-10f}h^{-3/2} b\,  b_T^{2}\right)   \ ,  \label{byxx} \\
\langle T^z{_z} \rangle & = & \frac{1}{2\kappa_5^2} r^5 \left( 1-\frac{r_h^4}{r^4} \right) \partial_r \log 
\left(e^{-10f}h^{-3/2} b^{2} b_T^{2}\right)   \ .  \label{byzz}
\eeqa
Expressions \eqref{bytt}-\eqref{byzz} can be expanded in powers of $\epsilon_h$ using the solution described in appendix \ref{blackholeapp}. With this we can write $\langle \tau^i{_j} \rangle =   \langle\tau^i{_j} \rangle_0 + \epsilon_h  \langle\tau^i{_j} \rangle_1 + \mathcal{O}(\epsilon_h^2)$. 

At zeroth order in $\epsilon_h$ we have $b=1$, since this function describes the anisotropy between the directions perpendicular to the magnetic field and the direction along the magnetic field, which is an order $\epsilon_h$ effect caused by the presence of fundamental matter. Therefore, at zeroth order in $\epsilon_h$ one obtains that $ \langle T^x{_x} \rangle =  \langle T^y{_y} \rangle = \langle T^z{_z} \rangle$ and the Brown-York tensor is isotropic. Actually, at zeroth order the solution to the type IIB action is nothing but $AdS_5\times S^5$ by construction, and we know already what the Brown-York tensor is going to be. The explicit calculation goes as
\be
\left( \sqrt{-\gamma} \,  \tau^i{_j} \right)_{0,div} = \frac{1}{ \kappa_{5}^2} (3r_\Lambda^4-r_h^4) \, \mathrm{diag} \left(  -3, 1,1,1  \right) \ ,
\ee
where the subindex $div$ signs that the expression is divergent in the $r_\Lambda\to \infty $ limit and must be regularized. Once again, the regularization is achieved by background subtraction 
\be
\langle T^i{_j} \rangle_0 = \lim_{r_\Lambda\to\infty} \left(  \left( \sqrt{-\gamma} \,  \tau^i{_j} \right)_{0,div} - \sqrt{1- \frac{r_h^4}{r_\Lambda^4}}  \lim_{r_h\to0 ,{B}\to0} \left( \sqrt{-\gamma} \,  \tau^i{_j} \right)_{0,div} \right) \ ,
\ee
 where the factor in the square root matches the euclidean geometries at the cutoff. A straightforward calculation gives
\be
\langle T^i{_j} \rangle_0 = \frac{V_{SE}}{2 \kappa_{10}^2} r_h^4 \, \mathrm{diag} \left(  -3, 1,1,1  \right) \ ,
\ee
after use of $\kappa_5^2= \kappa_{10}^2/V_{SE}$. Considering now the observation made in \eqref{tempentropy}, we can rewrite this expression as
\be
\langle T^i{_j} \rangle_0= \frac{sT}{4}  \mathrm{diag} \left(  -3, 1,1,1  \right) \ .
\ee
This, of course, is just the $AdS_5$ result, which gives an isotropic contribution. Notice that even when the entropy density and the temperature are sensitive to the magnetization of the plasma, their product cancels out factors coming from ${\cal M}$ to give the contribution to the SE tensor given above.

We consider now the contribution due to the presence of fundamental matter at first order in $\epsilon_h$, $\langle T^i{_j} \rangle_1 $. This term is given prior to regularization by
\be \label{epsilonBYcorrection}
\left( \sqrt{-\gamma} \,  \tau^i{_j} \right)_{1,div} = \frac{r_\Lambda(r_\Lambda^4-r_h^4)}{4 \kappa_5^2} \left[ \left( b_1'-3\Lambda_1'+\Upsilon_1' \right) \mathbb{I}_{4\times4}- 2 b_1'  \mathrm{diag} \left( 0,1,1,0 \right) \right]\ .
\ee
Once regularized we read the vev of the field theory SE tensor. In this case the $\langle T^t{_t} \rangle$ component must coincide with the ADM mass calculation (see appendix \ref{admandby}) given in the previous section. We have checked this explicitly by regularizing \eqref{epsilonBYcorrection} and evaluating the expression one gets in terms of $b_1$, $\Lambda_1$ and $\Upsilon_1$. This fact helps us to find the expressions for the SE tensor with the aim of the following two properties
 \beqa
\left( \sqrt{-\gamma} \, \tau^t{_t} \right)_{1,div} & = & \left( \sqrt{-\gamma} \,  \tau^z{_z} \right)_{1,div} \ , \label{eq.tau11}\\
\left( \sqrt{-\gamma} \,  \tau^x{_x} \right)_{1,div} & = &  \left( \sqrt{-\gamma} \,  \tau^y{_y} \right)_{1,div} = \left( \sqrt{-\gamma} \,  \tau^z{_z} \right)_{1,div} - \frac{r_\Lambda(r_\Lambda^4-r_h^4)}{2\kappa_5^2} b_1'  \ . \label{eq.tau12}
 \eeqa

Expression \eqref{eq.tau11} tells us that the contribution at first order in $\epsilon_h$ for $\langle T^t{_t} \rangle$ and $\langle T^z{_z} \rangle$ coincide, and since we know that the time component is given by the ADM energy, which we already calculated, we get
\be
\langle T^t{_t} \rangle = -\varepsilon_{ADM}=-\frac{3}{4}sT+\frac{1}{2} {B}{\cal M} \quad \Rightarrow \quad \langle T^z{_z} \rangle = \frac{1}{4}sT + \frac{1}{2} {B}{\cal M}=-{\cal G} \ ,
\ee
as announced.

To evaluate the pressure in the transverse directions $P_\perp=\langle T^x{_x} \rangle$ we can make use of the relation \eqref{eq.tau12}. There are two equivalent ways to obtain the answer. The first and more obvious one is to evaluate the $b_1'$ contribution in the r.h.s. of \eqref{eq.tau12} and regularize. This can be seen to lead to
\be
\langle T^x{_x} \rangle=\langle T^z{_z} \rangle-{B}{\cal M} = -{\cal F} \ ,
\ee
which is the expected result. Unfortunately, the evaluation makes use of the analytic --but somehow complicated-- form of $b_1(r)$, and intermediate steps to arrive to this result imply writing down long, non-illuminating expressions. A second strategy would be to notice that the contributions  to the regularized SE tensor from $\Lambda_1$ and $\Upsilon_1$ vanish. This implies that the correction at order $\epsilon_h$ to $\langle T^x{_x} \rangle$ is opposite in sign to the correction to $\langle T^z{_z} \rangle$, giving once again $\langle T^x{_x} \rangle= \frac{1}{4}sT - \frac{1}{2} {B}{\cal M}=-{\cal F} $. However, the explicit solution for $\Lambda_1$ and $\Upsilon_1$ is more complicated that the one for $b_1$, and intermediate expressions are again cumbersome equations which would lengthen this section without adding anything relevant.

Of course, from the former arguments it follows that the anisotropic measure is given by $\Delta_P={B}{\cal M}$, in agreement with the thermodynamic argument of the previous subsection.


\section{Energy loss in the magnetically anisotropic plasma}\label{sec:loss}

In this section we will focus on calculating the energy loss of the partons as they propagate in an anisotropic plasma. 

An estimation of this influence is coming through the calculation of the jet quenching parameter $\hat q$. In \cite{Liu:2006ug}, using the eikonal approximation in the 
high energy limit, they presented a non-perturbative prescription for calculating $\hat q$  as the coefficient of $L^2$ in an almost  light-like Wilson loop with dimensions $L^{-}\gg L$.
Following this prescription we will calculate $\hat q$ for our anisotropic backreacted background.

Another estimate of the energy loss of a parton passing through a medium may come through the drag force calculation. 
This computation can be implemented in a holographic framework through a macroscopic string moving with constant velocity $v$.
That string is attached to a probe flavor brane and dragged by a constant force $f$ which keeps the velocity fixed.
The drag coefficient $\mu$, measuring the energy loss, is calculated from the equation of motion 
\begin{equation}
f \, = \, \mu  \, p \ , 
\end{equation}
where $p$ is the parton momentum.
Following the prescription of \cite{Dragforce}, we will calculate $\mu$ for our anisotropic backreacted background.


\subsection{Drag force}

In this section we perform the calculation of a second observable
describing energy loss in the Quark-Gluon plasma by computing the drag
force experienced by an infinitely massive quark propagating at
constant velocity through an anisotropic plasma in constant magnetic field. 
In an anisotropic medium, the drag coefficient is not just a number
but  a matrix. This matrix is diagonal,  $\mu = \mbox{diag} (\mu_x,
\mu_y, \mu_z)$
but with  $\mu_x = \mu_y \neq \mu_z$, which means that the force and
the momentum (or the velocity) of the quark will not be aligned in
general.

The external magnetic field plays a double role in this scenario. On one side it makes the plasma anisotropic, on the other side it stimulates synchrotron radiation of gluons which is an additional factor contributing to the energy loss of the moving quark. In our analysis we stabilize the classical trajectory of the quark by introducing an additional electric field perpendicular to the magnetic field, compensating the Lorentz force. In addition we add a drag electric force compensating the viscous force of the plasma. The only energy loss is due to the negative work exerted by the viscous force. 

We follow closely \cite{Chernicoff:2012iq}, where the isotropic analysis of  \cite{Dragforce} has been generalized to the case of anisotropic plasma. Another relevant papers are ref. \cite{Kiritsis:2011ha}, where heavy quark in external magnetic field has been studied, and ref. \cite{Fadafan:2012qu} where the study of the radiation of a quark in an anisotropic plasma is performed.

On the gravity side the quark is described by a string propagating  in
the background \eqref{gen-metric} while the string action is given by
\begin{equation}
S \, = \, - \,  \frac{1}{2\pi\alpha'}\int d^2\sigma\,  \sqrt{-g}\,+\,  \frac{1}{2\pi\alpha'}\int P[B] \, =
\,  \int d^2\sigma \, {\cal L}  \ ,
\label{string-action}
\end{equation}
where $g$ is the induced worldsheet metric and $P[B]$ is the pullback of the Kalb-Rammond $B$-field. 

Physically, the electric forces needed to stabilize the trajectory are introduced by attaching one end of the string to a D7-brane and turning on a constant gauge filed $F_{MN}=\partial_{[M}A_{N]}$ on the brane \cite{Chernicoff:2012iq}. This results to the following boundary term
\begin{equation}
S_{\rm{bdry}}=\int_{\partial\Sigma}d\tau A_{N}\partial_{\tau}X^{N}=\frac{1}{2}\int_{\partial\Sigma}d\tau F_{MN}X^{M}\partial_{\tau}X^{N} \ .
\end{equation}
Next we define
\begin{equation}
\Pi_{M}=\frac{\partial {\cal L}}{\partial(\partial_{\sigma}X^{M})}\ .
\end{equation}
From the variation of the boundary action one has
\begin{equation}
\Pi_{M}|_{\partial\Sigma}+(2\pi\alpha')F_{MN}\partial_{\tau}X^{N}=0\ .\label{Newton}
\end{equation}
Equation (\ref{Newton}) is the equation for the balance of the forces acting on the moving quark. One can expect that the contribution to $\Pi_{M}$ from the Nambu-Goto term of the action corresponds to the viscous force of the plasma, while the contribution from the anti-symmetric part corresponds to the Lorentz force, this is confirmed by our analysis.

We choose an ansatz in which the string does not
move along the compact directions,
while due to the
rotational symmetry in the $xy$-plane, we can choose $y=0$. However in order for this ansatz to be consistent we need to compensate the Lorentz force along the $y$ direction. To clarify this we  keep a general ansatz $y=y(\tau,\sigma)$ for a while. We fix the reparameterization invariance by identifying $(t,r)=(\tau,\sigma)$
and consider a string profile of the form
\begin{equation}
x\, = \, \left[ u t \, + \, x(r) \right] \, \sin \varphi \ , \qquad z
\, = \,   \left[ u t \, + \,  z(r) \right] \, \cos \varphi \ , \quad y\,= \,y(\tau,\sigma)\ ,
\end{equation}
corresponding to a quark moving with velocity $u$ in the $xz$-plane at
an angle $\varphi$ with the $z$-axis and with so far undetermined profile along $y$ (eventually we will fix $y\equiv 0$).
Since the lagrangian does not depend explicitly on $x\,,\,y$ and $z$  we have:
\begin{eqnarray}
\label{conserved-pi_x}
\hspace{-1.7mm}\Pi_x &=& -\,  \frac{G_{11}}{\cal L} \,       \,\Big[ G_{tt} \,x'  \, +
\, u^2 \, \cos^2 \varphi \,\,  G_{33} \left(x' \, - \, z'\right)+G_{22}\dot y\left(\dot y x'-u y'\right) \Big]
\sin \varphi-B_{12}\dot y  \ ,
\\ [1.7mm]
\label{conserved-pi_y}
\Pi_y &=& - \,  \frac{G_{22}}{\cal L} \,\Big[ G_{tt} \,y'  \, +
\, u\, \sin^2 \varphi \,\,  G_{11} \left(uy' \, - \, x'\dot y\right)+u\, \cos^2\varphi \, G_{33}\left(u y'-z'\dot y\right)  \Big]
\\
&+&u\sin\varphi\,B_{12}\nonumber\\ [1.7mm]
\label{conserved-pi_z}
\Pi_z &=& - \,  \frac{1}{\cal L} \, G_{33} \,\Big[ G_{tt} \,z'  \, -
\, u^2 \, \sin^2 \varphi \,\,  G_{11} \left(x' \, - \, z'\right)+G_{22}\dot y\left(\dot y z'-u y'\right)  \Big]
\cos \varphi   \,,
\end{eqnarray}
where $'$ denotes differentiation with respect to $r$. As one can see from equation (\ref{conserved-pi_y}), even if we set $y\equiv 0$, $\Pi_y$ has a non-vanishing contribution $u\sin\varphi\,B_{12}$. For the $y$ component of equation (\ref{Newton}) we obtain
\begin{equation}
u\sin\varphi\,H_0-(2\pi\alpha')F_{02}=0\ ,\label{balance}
\end{equation}
where we have used that $F_{MN}$ has only electric components and that $B_{12}|_{\partial\Sigma}=H_{0}$. It is clear that equation (\ref{balance}) represents the balance between the Lorentz force $u\sin\varphi\,H_0$ acting on the quark and the electric field along the $y$ component, $F_{02}$, needed to cancel the Lorentz force. Therefore we fix $F_{02}=u\sin\varphi\,H_0/(2\pi\alpha')$ which enables us to set $y\equiv 0$.  The expressions for $\Pi_x$ and $\Pi_y$ simplify to:
\begin{eqnarray}
\label{conserved1-pi_x}
\Pi_x &=& -\,  \frac{1}{\cal L} \,      G_{11} \,\Big[ G_{tt} \,x'  \, +
\, u^2 \, \cos^2 \varphi \,\,  G_{33} \left(x' \, - \, z'\right) \Big]
\sin \varphi  \ ,
\\ [1.7mm]
\label{conserved1-pi_z}
\Pi_z &=& - \,  \frac{1}{\cal L} \, G_{33} \,\Big[ G_{tt} \,z'  \, -
\, u^2 \, \sin^2 \varphi \,\,  G_{11} \left(x' \, - \, z'\right) \Big]
\cos \varphi   \,,
\end{eqnarray}
Inverting \eqref{conserved1-pi_x} and  \eqref{conserved1-pi_z} we have
\begin{equation} \label{invert-conserved}
(x')^2 \, = \, \frac{G_{33} \, G_{rr}}{G_{11} \, G_{tt}} \, \frac{u^2
N_x^2}{D \, - \, N_x N_z}
\ , \qquad
(z')^2 \, = \, \frac{G_{11} \, G_{rr}}{G_{33} \, G_{tt}} \, \frac{u^2
N_z^2}{D \, - \, N_x N_z}\ ,
\end{equation}
with
\begin{eqnarray}
N_x&=& \Pi_x\, G_{tt} \, \cot \varphi  \, + \, u^2 \,  G_{11} \, \cos
\varphi \left(\Pi_z \, \cos \varphi \, + \, \Pi_x \, \sin \varphi
\right) \,, \\[1.9mm]
N_z&=& \Pi_z\, G_{tt} \, + \, u^2 \,  G_{33} \, \cos \varphi
\left(\Pi_z \, \cos \varphi \, + \, \Pi_x \, \sin \varphi \right)  \,,
\end{eqnarray}
and
\begin{eqnarray}
& D \, =\,  \cot \varphi \, G_{tt} \, D_1 \, D_2 &   \\[1.7mm]
&  D_1 \, = \, \Pi_x \Pi_z \, - \, \frac{1}{2} \, u^2 \, G_{11} \,
G_{33} \,  \sin 2 \varphi \quad  and \quad
D_2 \, = \, G_{tt} \, + \, u^2 \, \left( G_{33} \, \cos^2 \varphi \, +
\,  G_{11} \, \sin^2 \varphi \right) \ . &
\nonumber
\end{eqnarray}
The critical value of $r$ is determined from the equations
$
N_x\, =\, N_z \, = \, 0 $ and $D \, = \,N_x \, N_z \, = \, 0
$
and it is given by
\begin{equation}
r_c\, = \,  \frac{r_h}{\left(1 - u^2 \right)^{1/4}} \, \Bigg[1 \, + \,
\epsilon_h \, \frac{r_m^4}{r_h^4} \, \frac{u^2 \, \cos^2 \varphi}{1  -
u^2} \,
\left[\pi^2 \, + \, \log^2 \left(1 -  u^2 \right) \, + \,
\text{Li}_2\left(\frac{u^2}{u^2 - 1}\right)\right]\Bigg] \ .
\end{equation}
For this critical value we have
$\Pi_x =  u \, G_{11} \sin \varphi$ and $\Pi_z =  u \, G_{33} \cos \varphi$.
With all these constraints the denominator in \eqref{invert-conserved}
is always real and positive except at $r_c$, where it vanishes.
The numerators, at the critical point, also vanish and the functions
$x', z'$ are  smooth and negative for $r_h< r< r_s\to\infty$.
The force that must be exerted on the quark in order to maintain its
stationary motion (see \cite{Chernicoff:2012iq} for a detailed
explanation) is
\begin{equation}
\vec F \,  = \, \frac{1}{2 \pi \alpha'} \,  \left(\Pi_x, \Pi_z \right) \ ,
\end{equation}
in terms of the quark's velocity $\vec u=u(\sin \varphi, \cos
\varphi)$. For the specific case of anisotropic plasma under study we
have, in the small magnetic field limit
\begin{equation}
 F_{\{x,z\}} \,  = \, \frac{\pi}{2} \, \sqrt{\lambda_h} \, T^2
\frac{u}{\sqrt{1-u^2}} \Bigg[1 \, + \,\frac{1}{8} \, \epsilon_h \left[
2 \,- \,  \log (1-u^2) \, +  f_{\{x,z\}} \right] \Bigg]\ ,
\end{equation}
with
\begin{eqnarray}
f_x&=& \frac{1}{12}\, s_{\varphi} \,  \frac{r_m^4}{r_h^4} \, \Bigg[ 36
\, -12 \log \left(1 - u^2 \right) \, - \, 6
\left(1 \, + \frac{2}{1-u^2} \, - \,\frac{ 2 \,  c_{\varphi}^2 \,
u^2}{1-u^2} \right)  \times
\nonumber \\
&&
\quad \quad \quad \quad \quad \quad \quad \quad \quad \quad \quad \quad
\times \left[ \frac{\pi^2}{6} \, + \, \frac{1}{2}\log^2 \left(1 -  u^2
\right) \, + \,  \text{Li}_2\left(\frac{u^2}{u^2 - 1}\right)\right]
\Bigg]    \  , \\
f_z&=&  \frac{1}{12}\, c_{\varphi} \,  \frac{r_m^4}{r_h^4}  \, \Bigg[
36 \, -12 \log \left(1 - u^2 \right) \, + \, 6
\left(1 \,  - \,\frac{ 2 \,  s_{\varphi}^2 \, u^2}{1-u^2} \right)  \times
\nonumber \\
&&
\quad \quad \quad \quad \quad \quad \quad \quad \quad \quad \quad \quad
\times \left[ \frac{\pi^2}{6} \, + \, \frac{1}{2}\log^2 \left(1 -  u^2
\right) \, + \,  \text{Li}_2\left(\frac{u^2}{u^2 - 1}\right)\right]
\Bigg]   \ .
\end{eqnarray}
These expressions show the dependence of the magnitude of the force on
the magnetic field, and also how the presence of the anisotropy
affects the orientation of the field. This effect is \emph{not} due to
a Lorentz force, since it is compensated in equation (\ref{balance}). The directions
in which the quark is moving and the force is pushing are related by
\be
\arg \vec F = \varphi + \epsilon_h \frac{b_1(r_c)}{2} \sin (2\varphi) + {\cal
O}(\epsilon_h^2) \ ,
\ee
where $b_1(r_c)$ is the correction to the function $b$ in the ansatz
for the metric, evaluated at the critical radius.

Finally we would like to discuss the regime of validity of our analysis. In this section we introduce additional electric fields coupled to the charged quark, but we do not take into account the effect that they have on the SYM plasma. On the other hand we take into account the effect of the magnetic field on the plasma. This can be justified if the quark is moving sufficiently slow and the Lorentz force and the viscous force (both proportional to the velocity of the quark) can be compensated with electric fields weak relative to the magnetic field. Comparing to the discussion in section \ref{sec.regime},  for our solution to still  be reliable we consider that the electric field enters, effectively, at order $\epsilon^2$. This is the regime in which our analysis applies.


\subsection{Jet quenching}

In this section we will compute the jet quenching parameter $\hat q$
for our anisotropic plasma, extending the computation of
\cite{Bigazzi:2009bk}. 

We will follow the analysis of ref. \cite{Chernicoff:2012gu}, where jet quenching in anisotropic plasma have been studied. In our case the anisotropy is due to external magnetic field which triggers synchrotron radiation. Therefore one would expect the motion of the fundamental string to encode both the phenomena of jet quenching and the energy loss due to synchrotron radiation. This suggests that the prescription of refs. \cite{Liu:2006ug}, \cite{Chernicoff:2012gu} should be revised in the presence of external magnetic field. Technically this can be seen from the more general form of the fundamental string action (\ref{string-action}). In particular the presence of the $B$-field term affects the prescription for the holographic calculation of the wilson loop. More precisely  the on-shell action evaluated on the world sheet corresponding to the wilson loop is no longer purely imaginary which is crucial part of the derivations in refs. \cite{Liu:2006ug}, \cite{Chernicoff:2012gu}.

To circumvent this difficulty we will constraint ourselves to motion parallel to the magnetic field. In this case the pull-back of the $B$-field on the world sheet vanish and the motion of the string in not directly affected by the external magnetic field. Physically the Lorentz force acting on the quarks vanish. This suggests that in the holographic calculation the wilson loop is given by the area of the world sheet and technically there are no difficulties in applying the prescription of refs. \cite{Liu:2006ug}, \cite{Chernicoff:2012gu}. Note that the energy loss of the quark is still indirectly affected by the external magnetic field  through the alternated properties of the SYM plasma. 
We will sketch the derivation referring to \cite{Chernicoff:2012gu} for details\footnote{See \cite{Baier:2008js,Giataganas:2012zy}  as well.}.


We consider a parton moving along the direction parallel
to the magnetic field, $z$, with the momentum broadening taking place
in the transverse $xy$-plane.
Due to the rotational symmetry in the transverse plane we can choose
$\hat q$ to lay along the  $x$-direction.

In order to cover more general situations we consider the following
class of metrics
\begin{equation} \label{gen-metric}
ds^2_{10} \, = \, G_{tt}\, dt^2 \, + \, G_{11} \, dx^2 \,  + \, G_{22}
\, dy^2 \,  + \, G_{33} \, dz^2 \,  + \, G_{rr} \, dr^2 \, +   \cdots
\end{equation}
where the ellipses denote compact directions. After introducing
light-cone coordinates
\begin{equation}
z^\pm \, = \,  {\frac{1}{\sqrt{2}}} \, \left(t \, \pm \, z \right) \  ,
\end{equation}
we consider a rectangular Wilson loop with contour  $C$.
The expectation value of the Wilson loop is given by the extremum of
the Nambu-Goto action for a string with endpoints tracing the contour
$C$.
We consider a quark moving along $z^{-}$  and fix the
reparameterization invariance by identifying  $(z^-,r) \, = \,
(\tau,\sigma)$.
Moreover we set $z^{+}\,=\, 0$ and  specify the string embedding
through the function $x\,=\,x(r)$, subject to the boundary condition
$x(\pm \ell/2)=0$.
The Nambu-Goto action for this configuration is then given by
\begin{equation}  \label{trans-NG}
S\, = \, 2 i \, \frac{\sqrt{\lambda}}{2\pi R^2} \int{dz^{-}\int_0^{\ell/2}dr \,
\sqrt{\frac{1}{2} \left(G_{tt}\, + \, G_{33}\right) \left( G_{rr} \, +
\, G_{11} x'^2 \right)}} \ ,
\end{equation}
where $x'=dx/dr$ and the action is imaginary because the string
worldsheet is spacelike. Since the action does not depend on
$x$ explicitly we obtain the equation of motion for $x(r)$
\begin{equation} \label{EOMx}
x'^2 \, = \, \frac{\pi_x^2} {4 \, G_{11}\big(G_{tt} \, + \,
G_{33}\big) \, - \, \pi_x^2} \, \, \frac{G_{rr}}{G_{11}} \ .
\end{equation}
The turning point for the string is at $x'=0$ and following the
prescription of
\cite{Liu:2006ug} and \cite{Liu:2006he} we will work in the limit
$\ell \to 0$,  which corresponds to the limit  $\pi_x \to 0$.
Integrating \eqref{EOMx} and taking the limit for  $\pi_x \to 0$,  we
obtain the separation length between the endpoints of the string
\begin{equation} \label{trans-limit-sep}
L \, = \,  \pi_x {\cal I}_x + {\cal O}(\pi_x^2) \quad \text{with}
\quad {\cal I}_x \equiv \int^{r_*}_{r_h}
\frac{\sqrt{G_{rr}}}{G_{11}\, \sqrt{G_{tt}\, + \, G_{33}}}  \, dr \  .
\end{equation}
For the computation of the jet quenching parameter we have to evaluate
the action on shell and focus on the $L^2$ term after using
\eqref{trans-limit-sep}.
In way we have
\begin{equation}
S \, = \, \frac{i L^{-}}{8 \sqrt{2}} \, \frac{\sqrt{\lambda}}{2 \pi
R^2} \, \frac{2 L^2}{ {\cal I}_x} \ ,
\end{equation}
The prescription given in \cite{Liu:2006ug} and \cite{Liu:2006he} for
the jet quenching parameter is
\begin{equation}
e^{i 2 S} \, = \, \exp\left[
-\frac{L^{-}\ell^2}{4\sqrt{2}}\,\hat{q} \right] \, \quad \Rightarrow \quad
\hat{q} \, = \, \frac{\sqrt{\lambda}}{2 \pi R^2} \, \frac{2}{\mathcal{I}_x} \ .
\end{equation}
We rewrite this expression in terms of field theory quantities, namely
the 't Hooft coupling at the temperature scale, $\lambda_h \,
\equiv\lambda \,  e^{\Phi_h}$ and the temperature, given by
\eqref{temperature}. We obtain a correction to the unflavored
jet-quenching in the presence of a small magnetic field given by
\begin{equation}\label{qhat1}
\hat q = \frac{\pi^{3/2} \, \Gamma(\frac34)}{\Gamma(\frac54)} \,
\sqrt{\lambda_h}\, T^3
\Bigg[1 \, + \,  \frac{1}{8} \, \epsilon_h \, \left[ 2 \,+ \,  \pi \,
+ \, \frac{r_m^4}{r_h^4}
\left( 3 \, -\, \frac{1}{6} \, M_{cor} \right)  \, + \, {\cal O}
\left( \frac{r_m}{r_h}\right)^8 \right] \,
+ \,  {\cal O} (\epsilon_h^2)\Bigg] \ ,
\end{equation}
with
\begin{equation}
M_{cor} \, = \,  \pi \left(\log 8 - \pi \right) +  {}_4
F_3\left[1,1,1,\frac54;\frac74,2,2;1\right] \, \approx \, - 1.99143 \
.
\end{equation}
Unfortunately we were not able to find a closed expression for the jet
quenching parameter when generic magnetic field is present.




\subsection*{Comparison to the non-magnetic case}

To understand whether the presence of the magnetic field enhances or
reduces the energy loss parameterized by $\hat q$ with respect to a
theory without magnetic field, we must compare the expression
\eqref{qhat1} obtained before between the two
different theories.

The result in \eqref{qhat1} can be written as
\be
\hat q = \hat q_0 \left( 1+ \epsilon_h  \frac{r_m^4}{r_h^4} \cM +
\cO(\epsilon_h^2)  \right) \ ,
\ee
where
\be
\hat q_0 =  \frac{\pi^{3/2}  \Gamma(\frac34)}{\Gamma(\frac54)}
\sqrt{\lambda_h} T^3 \Bigg[ 1+\frac{\epsilon_h}{8} \left( 2+\pi
\right)+ \cO\left(\epsilon_h^2 \right)  \Bigg] \ ,
\ee
is the flavored result in the absence of a magnetic field
\cite{Bigazzi:2009bk}, which receives a enhancement of the jet
quenching with respect to the unflavored setup, both in a scheme where
the number of degrees of freedom (entropy density) and temperature are
kept fixed or in a scheme where the energy density and the force
between external quarks are fixed, indicating the robustness of the
correction. The presence of the magnetic field is given by the factor  $\cM=(3/8-M_{cor}/48)/8 \approx 5/12$ when the quark is moving along the direction of the transverse field.

To compare the magnetic and non-magnetic result we must state clearly
under which conditions we are making this comparison. For example, if
we choose to keep the entropy density and the temperature of the field
theories the same, therefore allowing us to compare the values of
$\hat q$ per degree of freedom at fixed $T$, we observe that the
parameters $N_c$  in the ${B}=0$ and ${B}\neq0$ cases are
related  (in the $r_m<r_h$ approximation) by
\be
N_{c,{B}} = N_{c,{B}=0} \Bigg[ 1- \frac{\epsilon_h}{4}
\frac{r_m^4}{r_h^4} + \cO \left( \epsilon_h^2 \right) \Bigg] \ .
\ee
This correction enters in the jet quenching parameter via
$\sqrt{\lambda_h} \sim N_c^{1/2}$, giving
\be
\frac{\hat q_{{B}}}{\hat q_{{B}=0}} = \Bigg[1+ \epsilon_h
\frac{r_m^4}{r_h^4} \left( \cM - \frac{1}{8} \right) +
\cO(\epsilon_h^2)  \Bigg] \ ,
\ee
implying that the presence of a magnetic field enhances the jet
quenching if the quarks are moving parallel to the magnetic field.

We could have chosen to fix $N_c$ as well as the entropy density,
allowing $T$ to vary. In that case the temperatures in the presence
and absence of a magnetic field are related by
\be
T_{{B}} = T_{{B}=0} \Bigg[1- \frac{\epsilon_h}{6}
\frac{r_m^4}{r_h^4} + \cO \left( \epsilon_h^2 \right) \Bigg]
\quad \Rightarrow \quad 
\frac{\hat q_{{B}}}{\hat q_{{B}=0}} =  \Bigg[1+
\epsilon_h  \frac{r_m^4}{r_h^4} \left( \cM - \frac{1}{2} \right) +
\cO(\epsilon_h^2)  \Bigg]  \ .
\ee
Therefore, in this scheme the presence of an anisotropy induced by the
magnetic field reduces the jet quenching for motion parallel to magnetic field.


\section{Conclusions}

In this work we have presented a solution to the equations of motion of type IIB supergravity in the presence of a smeared set of $N_f\gg1$ D7 branes. This solution is perturbative in the backreaction parameter $\epsilon_h\sim \lambda_h \frac{N_f}{N_c}$. The presence of a finite magnetic field sources an anisotropy in the solution which leaves a footprint in physical observables.

We have studied the thermodynamics associated to the magnetically anisotropic solution at first order in backreaction. Results for thermodynamic quantities like the entropy, free energy or magnetization coincide with studies performed in the quenched approximation, in which the dynamics of the matter in the fundamental representation decouples from the dynamics of the adjoint degrees of freedom. However, the quenched setup fails to describe the anisotropy due to the magnetic field. Actually, this probe approximation is only valid at small magnetic fields (compared to the scale of the temperature), where the anisotropy is very mild and the component of the NSNS potential, $B_{xy}(r)$, is approximately constant (see figure \ref{fig.h1}). At larger magnetic fields the backreaction of the branes onto the geometry creates a non trivial profile of the $B_{xy}$ component, which is itself reflected in an anisotropy in the metric.

We have presented expressions for the pressure of the plasma in the directions parallel and orthogonal to the magnetic field, that --not surprisingly-- do not coincide for finite magnetic field. This has as a consequence that the speed of sound in the two normal directions of the plasma do not coincide between them. Actually, from equations \eqref{cspara} and \eqref{csortho} we observe that both speeds of sound have a value lower than the conformal setup, signaling that scale invariance is broken by the magnetic field at first order in $\epsilon_h$, even when we have massless D7 branes (in the absence of magnetic field this is not the case, even when a charge density is present in the setup).

The breaking of the conformality means that we must not necessarily have a traceless stress-energy tensor. However, from direct inspection in section \ref{sec:setensor}, we have $\langle T^\mu{_\mu} \rangle=0$ at first order in backreaction. A word of caution is needed here. As discussed in some extent in the text, the $00$ component of the stress-energy tensor corresponds to the magnetic enthalpy, and not the internal energy of the system. Therefore one must not conclude that the tracelessness of the s-e tensor implies an equation of state of the form ${\cal U}=2 P_\perp+P_{||}$. In fact, given the relation between the magnetic enthalpy and the internal energy \eqref{internalenergy} we have
\begin{equation}
{\cal U} = 2 P_\perp+P_{||} + \, B {\cal M} \ .
\end{equation}

The last consequence of the anisotropy we have studied is the implications of the magnetic field in the energy loss of a heavy quark moving through the plasma. The lack of explicit isotropy has as a consequence that the energy loss depends on the direction of movement of the quark. Furthermore, if the quark is charged it will feel a Lorentz force due to the presence of the magnetic field.

An alternative approach to the jet quenching calculation we presented in this work is through  fluctuation analysis, relating the jet quenching parameter to  momentum broadening \cite{Gubser:2006nz}. In this reference the jet quenching is related to the transport coefficient associated to Langevin diffusion inside the plasma.
A similar analysis of the jet quenching parameter for a deformed ${\cal N}=4$ SYM after introducing massless flavor branes in the Veneziano limit was presented in \cite{Magana:2012kh}.
Since the background is analytic, but perturbative in the number of flavors, it is possible to obtain perturbative expressions for the jet quenching around the unquenched result. We
believe that the same analysis could be very well extended in our case, with the only concern on the complexity of the solution after the inclusion of the magnetic field. Generically, it seems a straightforward computation that will circumvent the issue of generalizing the "standard" recipe of the jet quenching calculation.

One interesting question that is raised immediately is what are the shear and bulk viscosities of the magnetic plasma. The shear viscosity is a tensorial quantity that has been seen in \cite{Rebhan:2011vd} not to satisfy the KSS value $\eta/s=1/4\pi$ for the anisotropic plasma of \cite{Mateos:2011ix} (this situation happens as well in condensed phases of holographic superfluids, as was proposed in \cite{Natsuume:2010ky} and checked explicitly in \cite{Erdmenger:2010xm}). Here we have an anisotropy sourced by a 2-form instead of a scalar, and this may complicate the analysis of the perturbations to calculate the shear viscosity via a Kubo formula. Due to the lack of conformality at first order in $\epsilon_h$ we expect that the bulk viscosity is non-zero as well, but proportional to $B {\cal M}$.


\section*{Acknowledgments}

We would like to thank Irene Amado,  Daniel Are\'an, Jorge Casalderrey-Solana, Aldo Cotrone and Mariano Chernicoff for comments.

The work of M.A. was supported by National Science Foundation grant PHY-07-57702.  The research of V.F. is supported by an IRCSET/Marie Curie fellowship. J.T. is supported by the Netherlands Organization for Scientific Research (NWO) under the FOM Foundation research program. D.~Z.~is funded by the FCT fellowship SFRH/BPD/62888/2009.
Centro de F\'{i}sica do Porto is partially funded by FCT through the projects PTDC/FIS/099293/2008 and CERN/FP/116358/2010.


\appendix


\section{Equations of motion from the effective action} \label{eff-EOM}

The expression for the one-dimensional effective lagrangian is given by ${\cal L}_{eff} $
\begin{eqnarray} \label{L-effective1}
{\cal L}_{eff} &=&
- \, \frac{1}{2} \left(\frac{h'}{h}\right)^2 \, + \,  12 \left(\frac{S'}{S}\right)^2 \,  + \, 8 \, \frac{F' S'}{F S} \, 
+ \, 24\, b_T^2\,b^2\, F^2\,S^6 - 4\, b_T^2\, b^2\, F^4\,S^4
\nonumber \\
&+&\left(\frac{b_T'}{b_T} \, + \, \frac{b'}{b}\right)\,\left( \frac{h'}{h}+ 8 \,\frac{S'}{S}+ 2\, \frac{F'}{F} \right) \, + \,  \frac{1}{2}\, 
\frac{b'}{b} \,  \left(\frac{b'}{b}+\frac{4b_T'}{b_T}\right) \, -  \, \frac{b_T^2\,b^2 Q_c^2}{2 h^2} 
\\
&-&\frac{1}{2}\,Q_f^2\, b_T^2\, b^2e^{2\Phi} S^8 \,
\left(1+\frac{e^{-\Phi}\,H^2\,h}{b^2}\right) \, - \, 4\,Q_f\, b_T^2\, b^2\,e^{\Phi}\,F^2\,S^6 \sqrt{1+\frac{e^{-\Phi}\,H^2\,h}{b^2}}
\nonumber \\ 
&-& \frac{1}{2}\,\Phi'^2
\, - \,  \frac{1}{2}\,\frac{e^{-\Phi}\,H'^2\,h}{b^2} \left(1 - \frac{e^{2\Phi}\,J'^2\,b^2}{b_T^2\, H'^{2}}\right) \, - \, Q_c H J' \ . \nonumber 
\end{eqnarray}
 Defining the following auxiliary (dimensionless) expressions 
\begin{equation}
\beta_1 \equiv \sqrt{1+\frac{e^{-\Phi}\,H^2\,h}{b^2}} \ , \quad
\beta_2 \equiv 1 + \frac{e^{2\Phi}\,J'^{2}\,b^2}{H'^2\, b_T^2} \quad and \quad
\beta_3 \equiv 1 + \frac{e^{-2\Phi}\,H'^{2}\,\beta_2}{Q_f^2\,H^2\, b_T^2\, b^2\,S^8}
\end{equation}
we can write the equations of motion in the following compact way
\begin{eqnarray}
\partial_\sigma^2(\log b_T)&=&0 
\label{diff-bT}\\ 
\partial_\sigma^2(\log b)&=&-\, \frac{4 Q_f\,H^2\, b_T^2\, h S^6 F^2}{\beta_1}
\,-\,e^{\Phi}\,H^2\,Q_f^2\, b_T^2\, h \,S^8\,\beta_3  \ , 
\label{diff-b} 
\end{eqnarray}
\begin{eqnarray}
\partial_\sigma^2(\log h)&=& \, - \, Q_c^2 \, \frac{b_T^2\, b^2}{h^2}\,-
\, \frac{2 Q_f\,H^2\, b_T^2\, h S^6 F^2}{\beta_1}\,-
\, \frac{1}{2}\,e^{\Phi}\,H^2\,Q_f^2\, b_T^2\,h \,S^8\,\beta_3 
\nonumber \\
&&\qquad \qquad \qquad  \qquad \quad \qquad \qquad  \quad \,\,\,\,
+\,\left(1-\beta_2\right)\,\frac{e^{-\Phi}\,h\,H'^2}{b^2} \ ,  \label{diff-h}
\end{eqnarray}
\begin{eqnarray}
\partial_\sigma^2(\log S)&=& - \, 2\, b_T^2\, b^2 F^4 S^4 \, + \,  6\, b_T^2\, b^2 F^2 S^6 
\, - \, \frac{Q_f \,e^\Phi\, b_T^2 \, b^2 F^2\,S^6}{\beta_1}
\nonumber \\
&&\qquad \qquad \qquad  \qquad \quad \qquad \qquad \,
+ \, \frac{1}{4}\,e^{\Phi}\,H^2\,Q_f^2\, b_T^2\,h \,S^8\, \beta_3 \ ,  \label{diff-S}
\end{eqnarray}
\begin{eqnarray}
\partial_\sigma^2(\log F)&=& 4\,b_T^2\, b^2 F^4 S^4 \, - 
\, \frac{1}{4}\,\left(1+\beta_1^2 \right)\,Q_f^2\,e^{2\Phi}\, b_T^2\, b^2\, S^8\,+
\, \frac{Q_f\,H^2\, b_T^2\, h S^6 F^2}{\beta_1}
\nonumber \\
&&\qquad \qquad \qquad  \qquad \quad \qquad \qquad \qquad \quad \qquad\,\,\,
+\,\frac{1}{4}\,\frac{e^{-\Phi}\,h\,H'^2\,\beta_2}{b^2} \ , \label{diff-F}
\end{eqnarray}
\begin{eqnarray}
\partial_\sigma^2\Phi&=& \frac{1}{2}\,\left(1+\beta_1^2 \right)
\Bigg[Q_f^2\,e^{2\Phi}\, b_T^2\,b^2\,S^8 \,+\, \frac{4 Q_f\, b_T^2\,b^2\, e^\Phi S^8}{\beta_1}\Bigg]\,
\nonumber \\
&&\qquad \qquad \qquad  \qquad \quad \qquad \qquad \qquad \quad \qquad\,\,\,
-\,\frac{1}{2}\,\frac{e^{-\Phi}\,h\,H'^2\,\beta_2}{b^2} \ , \label{diff-Phi}
\end{eqnarray}
\begin{eqnarray}
\partial_{\sigma} \left[\frac{e^{-\Phi}\,h\,H'}{b^2}\right]&=& e^{\Phi}\,Q_f^2\,H\, b_T^2\,h\,S^8\,+\,Q_c\,J'
+\, \frac{4 Q_f\,H\, b_T^2\, h S^6 F^2}{\beta_1} \ . \label{diff-H}
\end{eqnarray}
It is straightforward to check that the above set of equations, together with \eqref{defJ}, 
solve the full set of Einstein equations, provided the following ``zero-energy'' constraint is also satisfied
\begin{eqnarray} \label{constraint}
0&=&
-\frac{1}{2} \left(\frac{h'}{h}\right)^2 + 12 \left(\frac{S'}{S}\right)^2 + 8 \, \frac{F' S'}{F S}
- 24\, b_T^2\,b^2\, F^2\,S^6 + 4\, b_T^2\, b^2\, F^4\,S^4
\nonumber \\
&+&\left(\frac{b_T'}{b_T}+\frac{b'}{b}\right)\,\left( \frac{h'}{h}+ 8 \,\frac{S'}{S}+ 2\, \frac{F'}{F} \right)+  \frac{1}{2}\, 
\left(\frac{b'}{b}\right)\left(\frac{b'}{b}+\frac{4b_T'}{b_T}\right)+ \frac{b_T^2\,b^2 Q_c^2}{2 h^2} 
\\ \nonumber
&+&\frac{1}{2}\,Q_f^2\, b_T^2\, b^2e^{2\Phi} S^8 \,
\left(1+\frac{e^{-\Phi}\,H^2\,h}{b^2}\right)+4\,Q_f\, b_T^2\, b^2\,e^{\Phi}\,F^2\,S^6 \sqrt{1+\frac{e^{-\Phi}\,H^2\,h}{b^2}}
\\ \nonumber  \\ \nonumber
&-& \frac{1}{2}\,\Phi'^2
- \frac{1}{2}\,\frac{e^{-\Phi}\,H'^2\,h}{b^2} \left(1 - \frac{e^{2\Phi}\,J'^2\,b^2}{b_T^2\, H'^{2}}\right) \ .\nonumber 
\end{eqnarray}
This constraint can be thought of as the $\sigma\sigma$ component of the Einstein equations.
Differentiating \eqref{constraint} and using \eqref{defJ} and \eqref{diff-b}--\eqref{diff-H} we are getting zero,
meaning that the system is not overdetermined.


\section{Analytic perturbative solution of the equations of motion} \label{blackholeapp}

The homogeneous solutions for the equations \eqref{eomshot1}-\eqref{eomshot2} are 
\beqa
b_{1}^H & = & K_{b,1} + K_{b,2} \log\left[ 1-\frac{r_h^4}{r^4} \right]\ , \\ 
\Lambda_{1}^H & = & K_{\Lambda,1} \left(2\frac{r^4}{r_h^4}-1\right) + K_{\Lambda,2} \Bigg[2+\left(2\frac{r^4}{r_h^4}-1\right)\log\left[1-\frac{r_h^4}{r^4}\right]\Bigg] \ , \\
\Upsilon_{1}^H & = & K_{\Upsilon,1} \left(2\frac{r^4}{r_h^4}-1\right) + K_{\Upsilon,2} \Bigg[2+\left(2\frac{r^4}{r_h^4}-1\right) \log\left[1-\frac{r_h^4}{r^4}\right]\Bigg] \ , \\
\Delta_{1}^H & = & K_{\Delta,1} \,P_{1/2}\left(2\frac{r^4}{r_h^4}-1\right) + K_{\Delta,2} \,Q_{1/2}\left(2\frac{r^4}{r_h^4}-1\right) \ ,\\
\phi_{1}^H & = & K_{\phi,1} + K_{\phi,2} \log\left[ 1-\frac{r_h^4}{r^4} \right]\ , \\
H_{1}^H & = & K_{H,1} \, \frac{r^4}{r_h^4} + K_{H,2} \Bigg[ 1+ \frac{r^4}{r_h^4}  \log\left[ 1-\frac{r_h^4}{r^4} \right]  \Bigg] \ .
\eeqa
 The solutions with integration constant $K_{\Psi,1}$ are regular at the horizon whereas the ones with $K_{\Psi,2}$ diverge logarithmically there. This situation is reversed at infinity, where the solutions with $K_{\Psi,2}$ tend to zero whereas the ones with $K_{\Psi,1}$ can diverge or go to a constant. Also, in principle $Q_{1/2}(2r^4/r_h^4-1)$ has an imaginary part, but this is just $P_{1/2}(2r^4/r_h^4-1)$ and can be absorbed in $K_{\Delta,1}$. In the following, we will consider just the real part of $Q_{1/2}(2r^4/r_h^4-1)$ and that constants $K_{\Delta,1,2}$ are real.

The particular solution for every equation can be found in the following way. Defining
\begin{equation} \label{Gpsi-def}
G_\Psi(r) \equiv \frac{A_\Psi r^4+ B_\Psi r_m^4}{(r^4-r_h^4)\sqrt{r^4+r_m^4}} \ ,
\end{equation}
a particular solution of the corresponding inhomogeneous differential equation is given by
\begin{equation}
\Psi_1^{(p)}(r) \, = \, - \, \Psi_1^{(1)}(r) \int^r \frac{\Psi_1^{(2)}(\tilde r) \, G_\Psi(\tilde r)}{W(\tilde r)} \, d\tilde r \, 
+ \, \Psi_1^{(2)}(r) \int^r \frac{\Psi_1^{(1)}(\tilde r) \, G_\Psi(\tilde r)}{W(\tilde r)} \, d\tilde r \ ,
\end{equation}
with $\Psi_1^{(1,2)}$ the homogeneous solutions accompanied by the constants $K_{\Psi \, (1,2)}$ and 
\begin{equation}
W \, \equiv \, \Psi'_1{}^{(2)}(r)\Psi_1^{(1)}(r)-\Psi'_1{}^{(1)}(r)\Psi_1^{(2)}(r) \ , 
\end{equation}
the Wronskian. With the symbol $\int^r$ we denote an antiderivative, therefore no lower bound is considered 
(its addition amounts to a shift in the constants of integration for the homogeneous solutions). This method also works for $H_1$, but in that case
\begin{equation}
G_H(r) \equiv \frac{4r^4}{(r^4-r_h^4)\sqrt{r^4+r_m^4}}  \ .
\end{equation}
The solutions can be expressed in terms of the following expressions
\begin{equation}
ad_n(r) \, \equiv  \, \frac{1}{r_h^{n-1}} \int^r \frac{\tilde r^n\log\left[ 1-\frac{r_h^4}{\tilde r^4} \right]}{\sqrt{\tilde r^4 \, + \, r_m^4}} \,  d \tilde r \,  ,
\end{equation}
which for the special cases $n=1,\,5,\,9$ read
\begin{eqnarray}
ad_1(r)  &= & \frac{1}{4} \Bigg[   2 Li_2\left[ \frac{\alpha_r \, + \, 1}{\alpha_r \, - \, 1} \right] \,  - \,  
{ \rm Li}_2\left[ \frac{\alpha_r \, + \, 1}{\alpha_r \, - \, 1} \,  \frac{\alpha_{r_h} \, + \, 1}{\alpha_{r_h} \, - \, 1} \right] \,  - \, 
{ \rm Li}_2\left[ \frac{\alpha_r \, + \, 1}{\alpha_r \, - \, 1} \,  \frac{\alpha_{r_h} \, - \, 1}{\alpha_{r_h} \, + \, 1} \right] \Bigg] \,  ,
\nonumber \\
 ad_5 (r)& = & \, - \,  \frac{1}{2} \, \left( \alpha_{r_h}^2 \, - \, 1 \right)  ad_1(r) \, + \, \frac{1}{4} \, \alpha_{r} \, 
\frac{\alpha_{r_h}^2-1}{\alpha_{r}^2-1} \log \left[\frac{\alpha_{r_h}^2-\alpha_{r}^2}{\alpha_{r_h}^2-1}\right]  \, - \, 
\frac{1}{4} \, \log  \left[ \left( \alpha_{r_h}^2-1 \right) \, \frac{\alpha_{r} + 1}{\alpha_{r}-1}\right] \,
\nonumber \\
& + &  \,   
\frac{1}{4} \, \alpha_{r_h} \,  \log \left[\frac{\alpha_{r}  + \alpha_{r_h}}{\alpha_{r}  -  \alpha_{r_h}}\right] \ , 
\\
ad_9(r) & = & \frac{3}{8} \, \left( \alpha_{r_h}^2 \, - \, 1 \right)^2  ad_1(r) \, + \, 
\frac{1}{16} \, \alpha_{r} \, \left( 5 \, - \, 3  \alpha_{r}^2 \right) \, 
\left( \frac{\alpha_{r_h}^2-1}{\alpha_{r}^2-1} \right)^2 \,  \log \left[\frac{\alpha_{r_h}^2-\alpha_{r}^2}{\alpha_{r_h}^2-1}\right] \, - \, 
\frac{1}{8}  \alpha_{r} \,  \frac{\alpha_{r_h}^2-1}{\alpha_{r}^2-1}
\nonumber \\
& + & \frac{1}{16} \, \alpha_{r_h} \, \left( 5 \, - \, 3  \alpha_{r_h}^2 \right) \,  \log \left[\frac{\alpha_{r}+\alpha_{r_h}}{\alpha_{r}-\alpha_{r_h}}\right] \, - \, 
\frac{1}{4} \, \left(1 \, - \, \frac{1}{2} \, \alpha_{r_h}^2  \right) \,  \log  \left[ \left( \alpha_{r_h}^2-1 \right) \, \frac{\alpha_{r} + 1}{\alpha_{r}-1}\right] \ , 
\nonumber
\end{eqnarray}
with as usual 
\begin{equation}
\alpha_r \, \equiv \, \sqrt{1 \, + \, \frac{r_m^4}{r^4}}  \ , \quad 
\alpha_{r_h} \, \equiv \, \sqrt{1 \, + \, \frac{r_m^4}{r_h^4}} \ , \quad 
\alpha_{r_*} \, \equiv \, \sqrt{1 \, + \, \frac{r_m^4}{r_*^4}} \ .
\end{equation}
We express the solution\footnote{We do not include the expression for $\Delta_1$ since it is given in terms of integrals that cannot be evaluated analytically.
Furthermore, we do not need its explicit form in the text.}  
in terms of the dimensionless parameters $\alpha_r$,  $\alpha_{r_h}$ and $\alpha_{r_*}$.


\section*{Solution for $\phi_1$}

The solution for $\phi_1$ which is regular at the horizon and vanishes at $r\, = \, r_*$ is 
\begin{equation} \label{sol-phi1}
\phi_1   \, = \, \frac{1}{4} \log\left[\frac{\alpha_r + 1}{\alpha_r - 1} \, \frac{\alpha_{r_*} - 1}{\alpha_{r_*} + 1}\right] \, + \, 
\frac{1}{2} \, \alpha_{r_h} \, \log \left[ \frac{\alpha_{r_*} + \alpha_{r_h}}{\alpha_{r} + \alpha_{r_h}}\right] \ .
\end{equation}

Whenever we calculate a physical quantity we will set the scale $r_*$ at the horizon $r_*=r_h$, indicating that we are working in the IR range of energies.


\section*{Solution for $b_1$}

The solution for $b_1$ which is regular at the horizon and vanishes at $r\, = \, r_s$ is 
\begin{eqnarray} \label{sol-b1}
b_1 \,  & = &  \, \left( \alpha_{r_h}^2 - 1 \right) \Bigg[ \frac{1}{4} \log \left[ \frac{\alpha_{r_h}^2-\alpha_r^2}{\alpha_{r_h}^2-1} \right] 
\log \left[\frac{\alpha_r - 1}{\alpha_r + 1} \, \frac{\alpha_{r_h} + 1}{\alpha_{r_h} - 1}\right]  
\nonumber \\
&-& \frac{1}{4} \log \left[ \frac{\alpha_{r_h}^2-\alpha_{r_s}^2}{\alpha_{r_h}^2-1} \right] 
\log \left[\frac{\alpha_{r_s} - 1}{\alpha_{r_s} + 1} \, \frac{\alpha_{r_h} + 1}{\alpha_{r_h} - 1}\right] \, + \, ad_1(r) \, - \,  ad_1(r_s) \Bigg] \ .
\end{eqnarray}
The following limit of \eqref{sol-b1} will be useful in thermodynamic calculations
\begin{eqnarray} \label{limit-b1}
\lim_{rs \rightarrow \infty} b_1(r_h) \,  & = &  \, - \, \frac{1}{24} \, \left( \alpha_{r_h}^2 - 1 \right) \Bigg[ \pi^2 \,+ \, 12 \log^2 \left( \alpha_{r_h} \, - \, 1 \right)
\nonumber \\
& + &  \, 12 \log^2 \left( \alpha_{r_h} \, + \, 1 \right)  \,- \, 6 \log^2 \left( \alpha_{r_h}^2 \, - \, 1 \right) \, 
+ \, 12 \,  { \rm Li}_2\left[\frac{1 + \alpha_{r_h}}{1 - \alpha_{r_h} }  \right] \Bigg] \ .
\end{eqnarray}


\section*{Solution for $\Lambda_1$}

The solution for $\Lambda_1$\footnote{The value of the constant $ C_{\Lambda_1}$ is determined by requiring $\Lambda_1(r_s) \, = \, 0 $.} 
which is regular at the horizon and vanishes at $r\, = \, r_s$ is 
\begin{eqnarray}
\Lambda_1^{\rm{NH}} \,  & = &  \frac{1}{2} \, \frac{\alpha_{r_h}^2 \, - \, 1}{\alpha_r^2 \, - \, 1}
\left(2\alpha_{r_h}^2 \, - \, \alpha_r^2 \, - \, 1 \right)  
\left[ad_1(r) \,- \,  2 \,  ad_5(r) \, - \, \frac{1}{2} \,  \log  \left[ \left( \alpha_{r_h}^2-1 \right) \, \frac{\alpha_{r} + 1}{\alpha_{r}-1}\right]  \right]
\nonumber \\ 
&+& \frac{1}{4} \,\left( \alpha_{r_h}^2 \, - \, 1 \right)
\left(2 \, - \, \frac{\alpha_r^2  - 2\alpha_{r_h}^2 + 1}{\alpha_r^2 \, - \, 1}\log\left[\frac{\alpha_{r_h}^2-\alpha_r^2}{\alpha_{r_h}^2-1} \right]\right)
\\ 
&& \quad \quad \quad  \quad \quad \quad  \quad \quad \quad  \quad \quad 
\times \left( \alpha_{r} \,  \frac{\alpha_{r_h}^2 \, - \, 1}{\alpha_r^2 \, - \, 1} \, - \,  
\frac{1}{2} \, \alpha_{r_h}^2   \log  \left[ \left( \alpha_{r_h}^2-1 \right) \, \frac{\alpha_{r} + 1}{\alpha_{r}-1}\right]\right) \ , 
\nonumber
\end{eqnarray}
\begin{eqnarray}
\Lambda_1 \,  & = & C_{\Lambda_1} \, \frac{2\alpha_{r_h}^2 \, - \, \alpha_r^2 \, - \, 1}{\alpha_r^2 \, - \, 1} \, + \, \Lambda_1^{\rm{NH}}(r) 
\label{sol-L1} \\ 
&-& \frac{1}{4} \,\alpha_{r_h} \,\left( \alpha_{r_h}^2 \, - \, 1 \right) \left( 1 \, - \, \alpha_{r_h} 
\log \left[ 1 \, + \, \alpha_{r_h} \right]\right) 
\left(2 \, - \, \frac{\alpha_r^2  - 2\alpha_{r_h}^2 + 1}{\alpha_r^2 \, - \, 1}\log\left[\frac{\alpha_{r_h}^2-\alpha_r^2}{\alpha_{r_h}^2-1} \right]\right) \ .
\nonumber 
\end{eqnarray}
The following limit of \eqref{sol-L1} will be useful in thermodynamic calculations
\begin{eqnarray}
\lim_{rs \rightarrow \infty} \Lambda_1(r_h) \,  & = &  \, - \, \frac{1}{48} \, \left( \alpha_{r_h}^2 - 1 \right) \Bigg[12 \, + \,  \pi^2  \alpha_{r_h}^2
\, - \, 6 \, \alpha_{r_h}^2 \, \log^2 \left( \alpha_{r_h}^2 \, - \, 1 \right) 
\, + \, 12 \,  \alpha_{r_h}^2 \,  { \rm Li}_2\left[\frac{1 + \alpha_{r_h}}{1 - \alpha_{r_h} } \right]
\nonumber \\
&+& 24 \, \alpha_{r_h} \log \left[\frac{2  \alpha_{r_h}}{1 \, + \,  \alpha_{r_h}}\right] \, + \, 12  \alpha_{r_h}^2 \left( \log^2 \left[ \alpha_{r_h} \, - \, 1 \right] \, + 
\,  \log^2 \left[ \alpha_{r_h} \, + \, 1 \right]\right) \Bigg] \ . \label{limit-L1}
\end{eqnarray}


\section*{Solution for $\Upsilon_1$}

The solution for $\Upsilon_1$\footnote{The value of the constant $ C_{\Upsilon_1}$ is determined by requiring $\Upsilon_1(r_s) \, = \, {\frac{2}{9}} $.} is
\begin{eqnarray}
\Upsilon_1^{\rm{NH}} \,  & = & - \,  \frac{1}{2} \, \frac{2\alpha_{r_h}^2 \, - \, \alpha_r^2 \, - \, 1}{\alpha_r^2 \, - \, 1}
\Bigg[ \frac{1}{2}\, \left(\alpha_{r_h}^2 \, - \, 1 \right)  \log \left[\frac{ \alpha_{r} \, + \, 1}{\alpha_{r} \, - \, 1}\right] \, 
- \, 4 \, \alpha_{r} \, \frac{\alpha_{r_h}^2 \, - \, 1}{\alpha_r^2 \, - \, 1} 
\nonumber \\ 
&& \quad \quad \quad  \quad \quad \quad  \quad \quad \quad \quad 
+ \, 3 \, \left(\alpha_{r_h}^2 \, - \, 1 \right)  ad_1(r) \, + \,  2 \,  \left(7 \, - \, 3 \,  \alpha_{r_h}^2 \right)  ad_5(r) \, - \, 16 \, ad_9(r)   \Bigg]
\nonumber \\ 
&-& \frac{1}{4} \,\left( \alpha_{r_h}^2 \, - \, 1 \right)  \,\left( 3\,  \alpha_{r_h}^2 \, - \, 2 \right)
\left(2 \, - \, \frac{\alpha_r^2  - 2\alpha_{r_h}^2 + 1}{\alpha_r^2 \, - \, 1}\log\left[\frac{\alpha_{r_h}^2-\alpha_r^2}{\alpha_{r_h}^2-1} \right]\right)
\\ 
&&  \quad \quad \quad \quad \quad \quad  \quad \quad \quad \quad \quad
\times \left( \alpha_{r} \,  \frac{- \, 3 \, + \, 7 \, \alpha_{r_h}^2 \,  - \, \alpha_r^2  \left(1 \,  +  \, 3 \alpha^2_{r_h} \right)}{\left(\alpha_r^2 \, - \, 1 \right)^2 \, \left( 3\,  \alpha_{r_h}^2 \, - \, 2 \right)}
 \, + \,  \frac{1}{2} \,  \log  \left[ \frac{\alpha_{r} + 1}{\alpha_{r}-1}\right]\right) \ , 
\nonumber
\end{eqnarray}
\begin{eqnarray}
\Upsilon_1 \,  & = & C_{\Upsilon_1} \, \frac{2\alpha_{r_h}^2 \, - \, \alpha_r^2 \, - \, 1}{\alpha_r^2 \, - \, 1} \, + \, \Upsilon_1^{\rm{NH}}(r) \label{sol-Y1}
\\ 
&-& \frac{1}{8} \,\left( \alpha_{r_h}^2 \, - \, 1 \right) 
\left( 6 \alpha_{r_h} + (2 - 3 \alpha_{r_h}^2) \log \left[\frac{ \alpha_{r_h} \, + \, 1}{\alpha_{r_h} \, - \, 1}\right]  \right)
\left(2 \, - \, \frac{\alpha_r^2  - 2\alpha_{r_h}^2 + 1}{\alpha_r^2 \, - \, 1}\log\left[\frac{\alpha_{r_h}^2-\alpha_r^2}{\alpha_{r_h}^2-1} \right]\right) \ .
\nonumber 
\end{eqnarray}
The following limit of \eqref{sol-Y1} will be useful in thermodynamic calculations
\begin{eqnarray}
\lim_{rs \rightarrow \infty} \Upsilon_1(r_h) \,  & = & - \, \frac{1}{4} \Bigg[ 3 \, \alpha_{r_h}^2 \, - 4 \, \alpha_{r_h} \, -  \, 1 \, + \, 
6 \,\alpha_{r_h} \, \left(\alpha_{r_h}^2-1 \right) \log \left[\frac{2  \alpha_{r_h}}{1 \, + \,  \alpha_{r_h}}\right]
\\
&+&
\left(3 \alpha_{r_h}^2-2 \right) \, \left(\alpha_{r_h}^2-1 \right) 
\left( \frac{\pi^2}{12} \,+ \,  \text{Li}_2 \left[\frac{1 \, + \, \alpha_{r_h}}{1 \, + \, \alpha_{r_h}}  \right] \, 
+ \, \frac{1}{2} \, \log ^2\left[\frac{\alpha_{r_h} \, - \, 1}{\alpha_{r_h} \, + \, 1}\right] \right) \Bigg] \ . \label{limit-Y1}
\nonumber 
\end{eqnarray}


\section*{Solution for $H_1$}

The solution for $H_1$\footnote{The value of the constant $ C_{H_1}$ is determined by requiring $H_1(r_s) \, = \, 0 $.} is
\begin{eqnarray}
H_1^{\rm{NH}} \,  & = & - \,  \frac{\alpha_{r_h}^2 \, - \, 1}{\alpha_r^2 \, - \, 1} \, 
\left(  ad_5(r) \, + \, {\frac{1}{4}} \, \log \left[ \frac{\alpha_{r} + 1}{\alpha_{r}-1} \right]\right) 
\\ 
&+& {\frac{1}{4}} \left(\alpha_{r_h}^2 \, - \, 1\right)
\left(1 \, + \, \frac{\alpha_{r_h}^2 - 1}{\alpha_r^2 \, - \, 1}\log\left[\frac{\alpha_{r_h}^2-\alpha_r^2}{\alpha_{r_h}^2-1} \right]\right)
\left( \frac{\alpha_{r}}{ \alpha_{r}^2 - 1 } - \frac{1}{2} \log \left[\frac{ \alpha_{r} \, + \, 1}{\alpha_{r} \, - \, 1}\right]  \right) \ , 
\nonumber \\
H_1 \,  & = & C_{H_1} \,  \frac{\alpha_{r_h}^2 \, - \, 1}{\alpha_r^2 \, - \, 1} \, + \, H_1^{\rm{NH}}(r) \label{sol-H1}
\\ 
&-& \frac{1}{4} \, 
\left( \alpha_{r_h} + \frac{1}{2} (1 - \alpha_{r_h}^2) \log \left[\frac{ \alpha_{r_h} \, + \, 1}{\alpha_{r_h} \, - \, 1}\right]  \right)
\left(1 \, + \, \frac{\alpha_{r_h}^2 - 1}{\alpha_r^2 \, - \, 1}\log\left[\frac{\alpha_{r_h}^2-\alpha_r^2}{\alpha_{r_h}^2-1} \right]\right) \ .
\nonumber 
\end{eqnarray}
The limiting cases $r_h\to0$ and $r_m\to0$ can be obtained, recovering the results in \cite{Filev:2011mt} and \cite{Bigazzi:2009bk} respectively.


\section{Calculation of the Gibbs free energy} \label{Gpotential}

Applying the general recipe of \cite{Hawking:1995fd} we identify the on-shell Euclidean action, divided by the inverse temperature,  
with the Gibbs free energy (in an ensemble where the magnetic field is kept fixed). 
The Euclidean action has contributions from both bulk and surface terms given by the following expressions 
\begin{equation} 
{\cal I}_{\rm bulk} \, = \, \frac{V_{4}\pi^3}{2\kappa_{10}^2}\int {\cal L}_{\rm{II B}} \, d\sigma\  \quad and \quad 
{\cal I}_{\rm surf} \, = \, - \, \frac{V_{4}\pi^3}{\kappa_{10}^2}\sqrt{\gamma}K \ ,
\end{equation}
where ${\cal L}_{{\rm IIB}}$ is the wick rotated action \eqref{genact} and in ${\cal I}_{\rm surf}$ is the standard Gibbons-Hawking term. 
In order to generate  ${\cal L}_{eff}$, defined in \eqref{L-effective1}, we need to integrate by parts and get rid of second order derivative terms.
This in turn will lead us to consider boundary terms, on-shell, only of the form $\psi'_a / \psi_a$, where $\psi_a$ is a collective notation for the background functions $(b,b_T,h,F,S,\Phi)$.
While at $r\,=\,r_s$ those terms are canceled by the contribution of the Gibbons-Hawking term,  they remain at the horizon.  
It turns out that the only non-vanishing term is coming from the function $b_T$, with the following contribution
\begin{equation} 
- \, 2 \, \frac{\partial_{\sigma}b_T }{b_T} {\Bigg|}_{\sigma \, = \, \infty} = \, 4\, r_h^4 \ . 
\end{equation}
In other words 
\begin{equation} 
{\cal I} \, = \, {\cal I}_{\rm bulk}  \, + \, {\cal I}_{\rm surf} \, =  \, \beta \, \frac{\pi^3 V_3}{2 \, \kappa_{10}^2}  \,
\left( 4 \, r_h^4 \, - \,  \int {\cal L}_{\rm{eff}} \, d\sigma \right) \ . 
\end{equation}
Expanding in $\epsilon_h$ and using \eqref{expansion1} we have  

\begin{equation}
{\cal I}={\cal I}_0+\epsilon_h {\cal I}_{\rm DBI}+\epsilon_h {\cal I}_{\rm bound}+O(\epsilon_h^2)\ ,
\end{equation}
and more explicitly, changing the radial coordinate to $r$
\begin{eqnarray}
{\cal I}_0&=& \beta \, \frac{\pi^3 V_3}{2 \, \kappa_{10}^2}  \, \left( 6\, r_s^4 \, - \, 2 \, r_h^4 \right) \ , \quad 
{\cal I}_{DBI}=- \, \beta \, \frac{\pi^3 V_3}{2 \, \kappa_{10}^2}  \, 4 \, \int_{r_h}^{r_s} dr \, r \, \sqrt{r^4 \, + \, r_m^4}
\nonumber \\
{\cal I}_{bound}&=&- \, \beta \, \frac{\pi^3 V_3}{2 \, \kappa_{10}^2}  r_h^4  \Bigg[ b_1(r_h) \, - \, 3 \Lambda_1(r_h) \, + \, \Upsilon_1(r_h) 
\\ 
 && \qquad  \qquad \qquad \qquad 
- \, \left(1 \, - 2\,  \frac{r_s^4}{r_h^4}\right) \left[ 5 b_1(r_s) \, + \, 5 \Lambda_1 (r_s) \, -\, \Upsilon_1(r_s)\right] \Bigg] \ .   \nonumber 
\end{eqnarray}
This action is infinite and should be regularized by subtracting the zero temperature on-shell Euclidean action.  
We take $r_s$ as the radial cut-off for the integrals, such that the
finite and zero temperature geometries coincide. Since $g^T_{tt}(r_s)\neq g^0_{tt}(r_s)$, 
we rescale the Euclidean time of the zero temperature solution in the following way \cite{Bigazzi:2009bk}
\begin{equation}
\beta_0 \, = \, \beta \, \Bigg(1-\frac{r_h^4}{r_s^4}\Bigg)^{1/2} \ , 
\end{equation}
where $\beta$ is the period of the Euclidean time of the finite temperature solution. Doing so it is easy to prove that
\begin{equation} \label{renaction}
I \,  = \,  
- \, \frac{1}{8} \, \beta \,  N_c^2 \, \pi^2 \,  T^4  \left[ 1 \, + \, \epsilon_h \, \left( - \, \frac{1}{2} + \sqrt{1+\frac{r_m^4}{r_h^4}} + \frac{r_m^4}{r_h^4}\, 
\log \left[ \frac{r_h^2+\sqrt{r_h^4+r_m^4}}{r_m^2} \right] \right) + {\cal O}(\epsilon_h^2)  \right] \ , 
\end{equation}
and consequently \eqref{gibbsenergy}.


\section{ADM energy and Brown-York tensor} \label{admandby}

In this appendix we show that the definition of the ADM energy used in \eqref{ADMdef} coincides with the calculation of the $tt$ component of the boundary Brown-York tensor, given by $E_{BY}= -\int \dd^3 x \sqrt{-\gamma} k_i \xi_j \tau^{ij}$, where $k$ is the unit-norm vector orthogonal to the $t=constant$ surfaces, $\xi=\partial_t$ is a Killing vector and
\be
\tau^{ij} = \frac{2}{\sqrt{-\gamma}} \frac{\delta S_{5d}}{\delta \gamma_{ij}} 
\ee
the Brown-York boundary tensor. The calculation is done by noticing that in the boundary metric, $\gamma_{ij}$, $b_T(r_s)$ enters only in the $\gamma_{tt}$ component, then we can trade $\gamma_{tt}$ by $b_T$ in the calculation\footnote{Notice that we express the radial dependence of fields in terms of the radial coordinates $r$ or $\sigma$ indistinctively, depending on which one is more convenient for every step.}
\be
b_T(r_s) \frac{\delta S_{5d} }{\delta b_T(r_s)} = b_T(r_s)\frac{\delta S_{5d} }{\delta \gamma_{tt}} \frac{\delta \gamma_{tt} }{\delta b_T(r_s)}  = 2\gamma_{tt}  \frac{\delta S_{5d} }{\delta \gamma_{tt}}  \ .
\ee
On the other hand
\be
\frac{\delta S_{5d} }{\delta b_T(r_s)} =  - \frac{V_{3}V_{SE}}{2\kappa_{10}^2}\frac{\delta \phantom{A}}{\delta b_T(r_s)} \int^{\sigma_s} {\cal L}_{eff} d\sigma\ ,
\ee
where we are using \eqref{L-effective1} to define ${\cal L}_{eff}$. Using the equation of motion for $b_T$ we can express the on-shell value of ${\cal L}_{eff}$ as a total derivative and perform the integral and the variation
\be
\frac{\delta S_{5d} }{\delta b_T(r_s)} =   \frac{V_{3}V_{SE}}{2\kappa_{10}^2} \frac{\partial L_{eff}}{\partial b_T'}\Bigg|_{\sigma_s}  =  \frac{V_{3}V_{SE}}{2\kappa_{10}^2}\frac{2}{b_T(\sigma_s)} \partial_\sigma \log \sqrt{h b^2 b_T^2 F^2 S^8} \Big|_{\sigma_s}  \ .
\ee
Therefore, from the BY definition
\be
E_{BY}= - \int \dd^3 x \sqrt{-\gamma} k_i \xi_j \tau^{ij} = - b_T(r_s) \frac{\delta S_{5d} }{\delta b_T(r_s)} = - \frac{V_{3}V_{SE}}{\kappa_{10}^2} \partial_\sigma \log \sqrt{h b^2 b_T^2 F^2 S^8} \Big|_{\sigma_s}
\ee

Now, the ADM mass is defined in \eqref{ADMdef} as
\be
E_{ADM}=-\frac{V_3 V_{SE}}{\kappa_{10}^2} \sqrt{-g_{tt}}  \sqrt{g_8} K_T = -\frac{V_3 V_{SE}}{\kappa_{10}^2} \frac{\sqrt{-g_{tt}}}{\sqrt{g_{\sigma\sigma}}} \partial_\sigma \sqrt{h b^2 F^2 S^8}= E_{BY} \ .
\ee
Therefore, as far as we implement the same regularization procedure, these two quantities coincide and, as we have argued in the main text, can be identified with the magnetic enthalpy in the field theory side.


\end{document}